\documentclass[a4paper,11pt]{article}
\usepackage[utf8]{inputenc}
\usepackage[T1]{fontenc}
\usepackage{lmodern}
\usepackage{amsmath,amsthm,amsfonts,amssymb,bm}
\usepackage{mathtools}
\mathtoolsset{showonlyrefs}   
\usepackage{dsfont}			
\usepackage{enumitem}
\usepackage{graphicx}
\usepackage{booktabs}
\usepackage{upgreek}
\usepackage[authoryear]{natbib}
\usepackage[right=2.5cm,left=2.5cm,top=2cm,bottom=2cm]{geometry}
\usepackage[font={small,it}]{caption}	
\usepackage[colorlinks,linkcolor=blue,citecolor=blue,urlcolor=blue]{hyperref}
\usepackage{xcolor}
\usepackage[thinlines]{easytable}
\usepackage{multirow}
\usepackage{float}
\usepackage[scaled]{berasans}
\usepackage{subfig}

\usepackage{authblk}

\usepackage{array}
\newcolumntype{L}[1]{>{\raggedright\let\newline\\\arraybackslash\hspace{0pt}}m{#1}}
\newcolumntype{C}[1]{>{\centering\let\newline\\\arraybackslash\hspace{0pt}}m{#1}}
\newcolumntype{R}[1]{>{\raggedleft\let\newline\\\arraybackslash\hspace{0pt}}m{#1}}

\newcommand{\overbar}[1]{\mkern 1.5mu\overline{\mkern-2mu#1\mkern-2mu}\mkern 1.5mu}

\newcommand{\MU}{\bm{\mu}}
\newcommand{\SIG}{\bm{\Sigma}}
\newcommand{\X}{\mathbf{X}}
\newcommand{\x}{\mathbf{x}}
\newcommand{\Y}{\mathbf{Y}}
\newcommand{\y}{\mathbf{y}}
\newcommand{\W}{\mathbf{W}}
\newcommand{\D}{\mathbf{D}}

\newcommand{\A}{\mathbf{A}}
\newcommand{\OO}{\mathbf{O}}
\newcommand{\CC}{\mathbf{C}}

\newcommand{\sumN}{\sum_{i=1}^N}
\newcommand{\I}{\mathbb{I}}
\newcommand{\Gen}{\textsf{Gen}}
\newcommand{\Cor}{\textsf{Cor}}
\newcommand{\Noi}{\textsf{Noi}}
\newcommand{\vbar}{\,\lvert\,}
\newcommand{\yp}{\y^{\text{\tiny P}}}
\newcommand{\yq}{\y^{\text{\tiny Q}}}

\makeatletter
\def\blfootnote{\xdef\@thefnmark{}\@footnotetext}
\makeatother

\title{ \textsc{Unobserved Classes and Extra Variables in High-dimensional Discriminant Analysis} }

\author[1*]{\small Michael Fop}
\author[2]{Pierre-Alexandre Mattei}
\author[2]{Charles Bouveyron}
\author[1]{Thomas Brendan Murphy}

\affil[1]{\footnotesize\em School of Mathematics \& Statistics and Insight Research Centre, University College Dublin, Ireland}
\affil[2]{\footnotesize\em Universit\'{e} C\^{o}te d'Azur, Inria, CNRS, Laboratoire J.A. Dieudonn\'{e}, Maasai team, France}
\affil[*]{\footnotesize \texttt{michael.fop@ucd.ie}}

\date{}

\begin{document}
\maketitle

\blfootnote{The work of Fop M. and Murphy T. B. was supported by the Science Foundation Ireland funded Insight Research Centre (SFI/12/RC/2289\_P2). The work of Mattei P. A. and Bouveyron C. has been supported by the French government, through the 3IA C\^{o}te d’Azur Investment in the Future project managed by the National Research Agency (ANR) with the reference numbers ANR-19-P3IA-0002.}

\begin{abstract}
In supervised classification problems, the test set may contain data points belonging to classes not observed in the learning phase. Moreover, the same units in the test data may be measured on a set of additional variables recorded at a subsequent stage with respect to when the learning sample was collected. In this situation, the classifier built in the learning phase needs to adapt to handle potential unknown classes and the extra dimensions. We introduce a model-based discriminant approach, Dimension-Adaptive Mixture Discriminant Analysis (D-AMDA), which can detect unobserved classes and adapt to the increasing dimensionality. Model estimation is carried out via a full inductive approach based on an EM algorithm. The method is then embedded in a more general framework for adaptive variable selection and classification suitable for data of large dimensions. A simulation study and an artificial experiment related to classification of adulterated honey samples are used to validate the ability of the proposed framework to deal with complex situations.
\end{abstract}

\smallskip
{\small\noindent \textbf{Keywords:} Adaptive supervised classification, conditional estimation, model-based discriminant analysis, unobserved classes, variable selection.}


\section{Introduction}
\label{intro}
Standard supervised classification approaches assume that all existing classes in the data have been observed during the learning phase. However, in some cases there could be the possibility of having units in the test set belonging to classes not previously observed. In such situation, a standard classifier would fail to detect the novel classes and would assign the observations only to the classes it is aware of from the learning stage. Moreover, the observations to be classified may be recorded on a collection of additional variables other than the variables already observed in the learning data. An example is classification of spectrometry data, where the test data may be measured at a finer resolution than the learning set, hence with a increased number of wavelengths. In this setting, the classifier would also need to adapt to the increasing dimensionality. The combination of these cases together leads to a complex situation where the model built in the learning stage is faced with two sources of criticality when classifying the new data: unobserved classes and extra variables.

In a recent work, \cite{bouveyron:2014} introduced an adaptive method for model-based classification when the test data contains unknown classes. Nonetheless, the method is not capable of handling the situation of additional variables. To deal with this problem, this work introduces a model-based adaptive classification method for detection of novel classes in a set of new data that is characterized by an expanded number of variables with respect to the learning set. The approach is developed in conjunction with an adaptive variable selection procedure used to select the variables of the test set most relevant for the classification of the observations into observed and novel classes. An EM algorithm based on an inductive approach is proposed for estimation of the model. Variable selection is performed with a greedy forward algorithm that exploits the inductive characteristics of the approach and make it suitable for high-dimensional data.

The methodology presented here aims at tackling the problems arising from a mismatch in the distributions of labels and input variables in training and test data. This problem is more generally denoted as "dataset shift", and we point the interested reader to \cite{quionero:2009} and \cite{moreno:2012}. In this work, the mismatch is due to unrepresented classes in the training data and increased dimensions of the test data. 

\subsection{Model-based discriminant analysis}
\label{da} 
Consider a set of learning observations $\{\x_s; \bar{\bm{\ell}}_{s}\}$, where $\x_s$ is the observation of a vector of random variables and $\bar{\bm{\ell}}_{s}$ is the associate class label, such that $\bar{\ell}_{sc} = 1$ if observation $s$ belongs to class $c$, 0 otherwise; $c = 1,\,\dots,\,C$. The aim of supervised classification is to build a classifier from the complete learning data $\lbrace \x_s, \bar{\bm{\ell}}_s \rbrace$ and use it to assign a new observation to one of the known classes. \emph{Model-based discriminant analysis} \citep[MDA,][]{bouveyron2019,mclachlan:2012,mclachlan:2004,fraley:raftery:2002} is a probabilistic approach for supervised classification of continuous data in which the data generating process is represented as follows:

\begin{equation}\label{eq:1}
   \begin{aligned}
    \bar{\bm{\ell}}_{s} &\sim \prod_{c=1}^C \tau_c^{ \,\bar{\ell}_{sc} },\\
    (\x_s \vbar \bar{\ell}_{sc} = 1) &\sim \mathcal{N}( \MU_c, \SIG_c ),
   \end{aligned}
\end{equation}
where 
$\tau_c$ denotes the probability of observing class $c$, with $\sum_c \tau_c = 1$. 
Consequently, the marginal density of each data point corresponds to the density of a Gaussian mixture distribution:
$$
f(\x_s \,; \bm{\Theta} ) = \sum_{c=1}^C \tau_c \, \phi ( \x_s \,;\, \MU_c, \SIG_c), 
$$
where $\phi( . \,;\, \MU_c, \SIG_c)$ is the multivariate Gaussian density, with mean $\MU_c$ and covariance matrix $\SIG_c$, and $\bm{\Theta}$ is the collection of all mixture parameters.
Then, using the \emph{maximum a posteriori} (MAP) rule, a new observation $\y_i$ is assigned to the class $\ell_{ic}$ with the highest posterior probability:
\begin{equation}\label{eq:map}
\Pr(\ell_{ic} = 1 \vbar \y_i) = 
\dfrac{ \tau_c \, \phi( \y_i \,;\, \MU_c, \SIG_c ) }{ \sum_{c=1}^C \tau_c \, \phi( \y_i \,;\, \MU_c, \SIG_c) }.
\end{equation}

The framework is closely related to other discriminant analysis methods. If the covariance matrices are constrained to be the same across the classes, then the standard linear discriminant analysis (LDA) is recovered. On the other hand, if the covariance matrices have no constraints, the method corresponds to the standard quadratic discriminant analysis \citep[QDA][]{mclachlan:2004,fraley:raftery:2002}. Several extension of this framework have been proposed in the literature in order to increase its flexibility and scope. For example, \cite{hastie:1996} consider the case where each class density is itself a mixture of Gaussian distributions with common covariance matrix and known number of components. \cite{fraley:raftery:2002} further generalize this approach, allowing the covariance matrices to be different across the sub-groups and applying model-based clustering to the observations of each class. Another approach, eigenvalue decomposition discriminant analysis  (EDDA, \citealp{bensmail:celeux:1996}), is based on the family of parsimonious Gaussian models of \cite{celeux:govaert:1995}, which imposes cross-constraints on the eigen-decomposition of the class covariance matrices. This latter approach allows more flexibility than LDA, and is more structured than QDA and the methods of \cite{fraley:raftery:2002}, which could be over-parameterized. In high-dimensional settings, different approaches have been proposed based on regularization and variable selection: \cite{friedman:1989} and \cite{xu:2009} propose regularized versions of discriminant analysis where a shrinkage parameter is introduced to control the degree of regularization between LDA and QDA; \cite{le:2020} and \cite{sun:2015} define frameworks where a penalty term is introduced and the classes are characterized by sparse inverse covariance matrices. It is also worth to mention that for high-dimensional data, the framework of discriminant analysis has often been phrased in terms of sparse discriminant vectors, see for example: \cite{clemmensen:2011}, \cite{mai:2012}, \cite{safo:2016}, \cite{jiang:2018}, \cite{qin:2018}.

\subsection{Adaptive mixture discriminant analysis}
\label{adapt}
The discriminant analysis approaches pointed out earlier assume that all existing classes have been observed in the training set during the learning phase, not taking into account that the test data might include observations arising from classes present in the learning phase. Initial works in the the context of unobserved classes detection and model-based discriminant analysis are those of \cite{miller:2003} and \cite{frame:2007}, while examples of applications include galaxy classification \citep{bazell:2005} and acoustic species classification \citep{woillez:2012}. More recently, building on \cite{miller:2003} work, \cite{bouveyron:2014} introduced \emph{Adaptive Mixture Discriminant Analysis} (AMDA), a framework for model-based discriminant analysis which allows the modeling of data where the test set contains novel classes not observed in the learning phase. The AMDA model considers the data arising from a mixture model with observed and unobserved classes and \cite{bouveyron:2014} proposes two alternative approaches for model estimation. In particular, the {\em inductive} approach, where the classifying function is first estimated on the learning set and then applied to the test data. Crucially, the core assumption of the inductive approach is that the parameters estimated on the training data are fixed when dealing with the test set \citep[see][for example]{chapelle:2006,pang:kasabov:2004}. The assumption makes the approach most suitable for fast on-line data classification when the data come in multiple streams. In fact, with this approach, the learning set does not need to be kept in memory for prediction on a set of new data points, only the estimated parameters need to be stored.

In what follows we provide a formal description of the problem of unobserved classes in the test data and give a brief overview of the inductive AMDA methodology, as it constitutes the starting block of the main contribution of this paper.
The learning data is composed of $M$ observations $\x_s$ and the associated class labels $\bar{\bm{\ell}}_s$, while the test data contains $N$ new observations $\y_i$. For ease of presentation, we treat the classes as sets, with ${\cal C}$ the set of all classes. The AMDA framework considers the situation where the data generating process is the same as depicted in \eqref{eq:1} but only a subset of classes is observed in the training set, that is a subset $\mathcal{K} \subseteq \mathcal{C}$ of classes has been represented in the learning data. Therefore the test data may contain a set of extra ``hidden'' classes $\mathcal{H}$ such that $\mathcal{K} \cup \mathcal{H} = \mathcal{C}$. The cardinality of these sets (i.e. the number of classes) is denoted with $K$, $H$ and $C$ respectively, such as $K + H = C$. 

The inductive AMDA approach consists of two phases: a learning phase and a discovery phase.
The initial {\em learning phase} corresponds to the estimation of a model-based discriminant analysis classifier using the training data. The data in the learning phase are complete, and the parameters estimated in this stage are then employed in the subsequent discovery phase. The {\em discovery phase} searches for $H$ novel classes in the set of new observations $\y_i$. In this phase, because of the inductive approach, the learning data is no longer needed and is discarded. The only relevant quantities to be retained are the parameter estimates obtained during the learning phase. In this stage, one needs to estimate the parameters of the unobserved classes in a partially unsupervised way in order to derive the classification rule as in \eqref{eq:map}. Because the observations $\y_i$ are unlabelled, the following log-likelihood is considered:
$$
L( \Y ; {\bm{\Theta}} ) = \sumN \text{log} \left\lbrace \sum_{k=1}^K \tau_k \, \phi ( \y_i \,;\, \overbar{\MU}_k, \overbar{\SIG}_k  ) + 
\sum_{h=K+1}^C \tau_h \, \phi \left( \y_i \,;\, \MU_h, \SIG_h  \right) \right\rbrace,
$$
where $\bm{\Theta}$ denotes the collection of all parameters. The parameters $\overbar{\MU}_k, \overbar{\SIG}_k$ for $k=1,\,\dots,\,K$ are those of the classes observed in the training set and have been already estimated in the learning phase; the bar in the notation indicates that at this stage these parameters have already been estimated and are fixed. On the other hand, the Gaussian density parameters $\MU_h, \SIG_h$ for $h=K+1,\,\dots,\,C$  remain to be estimated. Note that quantities related to the known classes are denoted with subscript $k$, while the subscript $h$ denotes quantities related to the new classes; subscript $c$ denotes both known and unknown classes. \cite{bouveyron:2014} presents an EM algorithm \citep{dempster:etal:1977,mclachlan:krishnan:2008} for optimization of the above log-likelihood with respect to the parameters of the unobserved classes, keeping fixed the parameters estimated in the learning phase.

\subsection{Contribution and organization of the paper}
\label{organization}
The present paper extends the inductive AMDA framework to the case where the test data includes not only unobserved classes, but also extra variables. The contribution of this work is twofold. First, we propose a novel inductive model-based adaptive classification framework which can model the situation where the observations of the test data may contain classes unobserved during the training stage and are recorded on an expanded set of input features. Secondly, we incorporate this framework in a computationally efficient inductive variable selection procedure employed to detect the most relevant variables for classification into observed and unknown classes.

The paper is organized as follows. The current section~\ref{intro} introduced the problem of classifcation with unknown classes and extra variables, also providing a short overview of model-based classification via discriminant analysis. In particular, section~\ref{adapt} briefly described the adaptive discriminant analysis method which is the basis of our proposed method. The following sections presents the novel methodology. Section~\ref{stream_adpat} introduces the novel adaptive mixture discriminant analysis method capable to handle a the complex situation where the new observations include information about unknown classes and are also measured on a set of additional variables. In Section~\ref{varsel}, the proposed method is naturally incorporated in a variable selection approach tailored for classification of high-dimensional data. Extensive simulation experiments are conducted in Section~\ref{sim}, in order to evaluate the performance of the proposed method for adaptive classification and variable selection. Section~\ref{appl} presents an application to the classification of spectroscopy data of contaminated honey samples. The paper ends with a discussion in Section~\ref{disc}.

\section{Dimension-adaptive mixture discriminant analysis}
\label{stream_adpat}
The AMDA framework combines supervised and unsupervised learning for detecting unobserved classes in the test data. 
However, in a dynamic classification setting, the new observations could be characterized not only by information about novel classes. In fact, the units in the test data could also have been recorded on a set of additional variables other than the ones already observed in the learning data. An example would be spectrometry data where the samples in the learning phase have been recorded over a set of specified wavelengths, and then additional samples have been collected in a subsequent phase at a finer resolution. Another example is the case where the variables correspond to points in time, and observations are recorded in a continuous manner; a given set of observations could have been collected up to a certain data point, while another set of units could have been recorded up to a successive period of time.  A further example is the case where some of the variables in the training data are corrupted and cannot be used to build the classifier, while they are available in the testing stage. 

Formally, we describe the setting of unknown classes and extra variables as follows. The learning data $\X$ is composed of $M$ observations $\x_s$ with the associated class labels $\bar{\bm{\ell}}_s$, and the test data $\Y$ is composed of $N$ new unlabelled observations $\y_i$. As in Section \ref{adapt}, the test data may contain a set of unobserved classes $\mathcal{H}$ such that $\mathcal{K} \cup \mathcal{H} = \mathcal{C}$.
However, in this setting we consider the case where only a subset of variables available in the test data are observed or recorded in the training data. Hence, the test data also includes extra variables compared to the data set used for training. 
We consider the collection of variables observed in learning and test data as sets. In the case where only a subset of variables is available in the learning set, the test observations $\y_i$ are realizations of the set of variables ${\cal R}$, while the training observations are recorded on the subset of variables ${\cal P}\subset {\cal R}$. Consequently, the set ${\cal Q} = {\cal R} \setminus {\cal P}$ denotes the set of additional variables observed in the test set but not in the training set. The cardinalities of these sets, i.e. the number of variables in each set, are indicated with $P$, $Q$, and $R$, respectively, with $R = P+Q$. 

The extra dimensions in the test data induce an augmented parameter space in the prediction and novel class detection stage of the classifier. Discarding the additional dimensions available in the test data can potentially damage the classification performance of the model, especially if the extra variables contain useful discriminant information. In this context, the classifier built in the learning phase needs to adapt in order to handle the situation where the new data to be classified contains information about novel classes and extra variables. To the purpose, we introduce \emph{Dimension-Adaptive Mixture Discriminant Analysis} (D-AMDA). The model is a generalization of AMDA and is designed to classify new observations measured on additional variables and possibly containing information about unobserved classes. Under the model, the joint densities of each observed and new data point together with observed and unobserved class labels are given by:
\begin{equation}
\label{eq:joint_train} 
 f(\x_s, \bar{\bm{\ell}}_s\,; \bm{\Theta}_x ) = \prod_{k=1}^K \left \lbrace \tau_k\,\phi(\x_s \,; \MU_k, \SIG_k) \right \rbrace^{\bar{\ell}_{sk}},
\end{equation}
\begin{equation}
\label{eq:joint_test}
f(\y_i, \bm{\ell}_i \,; \bm{\Theta}_y ) = \biggl[ \, \prod_{k=1}^K \left\lbrace \tau_k\,\phi(\y_i \,; \MU^*_k, \SIG^*_k) \right\rbrace^{\ell_{ik}} \biggr] \times \biggl[ \, \prod_{h=K+1}^C \left\lbrace \tau_h\,\phi(\y_i \,; \MU^*_h, \SIG^*_h) \right\rbrace^{\ell_{ih}} \biggr] ,
\end{equation}
with $s=1,\,\dots,\,M$, $i=1,\,\dots,\,N$, and $\bm{\Theta}_x$ and $\bm{\Theta}_y$ are the set of parameters for training and test observations. As earlier, the subscript $k$ indicates quantities related to the known classes, while the subscript $h$ denotes quantities related to the new classes.
The parameters $\MU_k$ and $\SIG_k$ are the class-specific mean and covariance parameters of the observed classes in the learning data and related to the subset of variables ${\cal P}$. The parameters denoted with $\MU^*$ and $\SIG^*$ denotes respectively the class-specific mean and covariance parameters for both observed and unobserved classes and related to the full collection of variables of the test data. These parameters are defined on an augmented space compared to the parameters in \eqref{eq:joint_train}. Indeed, $\MU_k$ and $\SIG_k$ are $P$-dimensional vectors and $P \times P$ matrices, while $\MU^*$ and $\SIG^*$ are $R$-dimensional vectors and $R \times R$ matrices. As such, the model takes into account that $\y_i$ may be measured on additional variables and generalizes the AMDA framework.

\begin{figure}[tb]
 \centering
 \subfloat[][\em Learning phase.]
 {\includegraphics[scale=1.3]{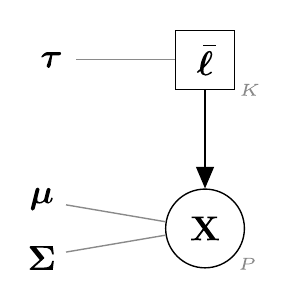}}\qquad\qquad
 \subfloat[][\em Discovery phase.]
 {\includegraphics[scale=1.3]{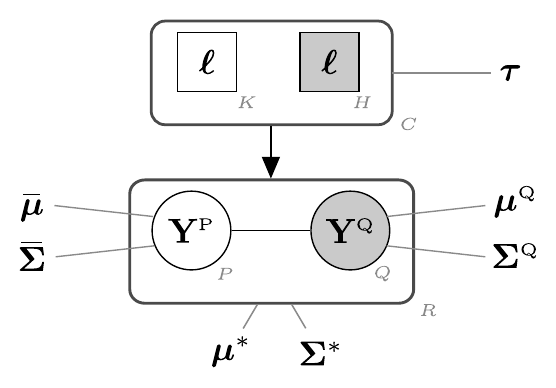}}
 \caption{\label{fig:frame}General framework of the inductive estimation approach for Dimension-Adaptive Mixture Discriminant Analysis.}
\end{figure}

Similarly to AMDA, model estimation for D-AMDA is carried out within an inductive estimation framework. Figure~\ref{fig:frame} provides a sketch of the general framework. In (a) the training data $\X$ and the corresponding collection of labels $\bar{\bm{\ell}}$ are observed. The aim of the learning stage is to estimate the set of parameters $\MU_k$, $\SIG_k$ and $\bm{\tau}$ of the density in \eqref{eq:joint_train}. In (b) only $\Y$ is observed and no information about the classification is given. The test data is partitioned into two parts: $\Y^{\text{\tiny P}}$, the subset of data corresponding to the variables observed in the training set, and $\Y^{\text{\tiny Q}}$, the subset of data related to the additional variables (gray background). In the discovery phase the aim is to estimate the parameters $\MU_k^*$, $\MU_h^*$, $\SIG_k^*$, $\SIG^*_h$ and $\bm{\tau}$ of the distribution in \eqref{eq:joint_test}, as well as to infer the classification of the new unlabelled data points. The collection of class labels to be inferred here is composed of the labels indicating the classes observed in the learning stage and the labels indicating the new classes (gray background). Model estimation for D-AMDA is detailed in the next sections.

\section{Inductive model estimation and inference}
The use of an inductive estimation approach is appropriate for the proposed D-AMDA framework, as it allows to only retain the test data once a set of mean and covariance parameter estimates have been obtained from the training data. Since the training set is lower dimensional compared to the test data, this can be particularly efficient in high-dimensional and on-line classification settings. Like in Section \ref{adapt}, the approach is composed of a learning and a discovery phase. In this case, the discovery phase includes a novel estimation procedure employed to account for the extra dimensions.

\subsection{Learning phase}
\label{learn}
The learning phase of the inductive approach consists of estimating parameters for the observed classes employing only the training set. From equation $\eqref{eq:joint_train}$, this stage  corresponds to the standard estimation of a model-based discriminant analysis classifier, performed by optimization of the associated log-likelihood:
$$
L(\X, \bar{\bm{\ell}}\,; \bm{\Theta}_x ) = \sum_{s=1}^M\sum_{k=1}^K {\bar{\ell}_{sk}} \log \left \lbrace \tau_k\,\phi(\x_s \,; \MU_k, \SIG_k) \right \rbrace,
$$
which reduces to the separate estimation of the class density parameters.
Here we consider the eigenvalue decomposition discriminant analysis (EDDA) of \cite{bensmail:celeux:1996}, in which the class covariance matrices $\SIG_k$ are parameterized according to the eigen-decomposition $\SIG_k = \lambda_k\,\D_k\,\A_k\,\D_k^{'}$, providing a collection of parsimonious models. Estimation of this model is carried out using the \texttt{mclust} R package \cite{scrucca:fop:2016}, which also automatically selects the best covariance decomposition model using the \emph{Bayesian Information Criterion} \citep[BIC,][]{schwarz:1978,fraley:raftery:2002}. This learning phase is more general than the one in \cite{bouveyron:2014}. In fact, the author considers a QDA model in the learning phase, a particular case of EDDA corresponding to an unconstrained covariance model \citep[see][]{scrucca:fop:2016}. The EDDA classifier learned in this phase is more flexible and is proven to perform better than QDA \citep{bensmail:celeux:1996}, although it will introduce some complications, which described in the following section.

The learning phase outputs the parameters of the EDDA model fitted on the training data $\overbar{\MU}_k$ and $\overbar{\SIG}_k$ for $k = 1, \ldots, K$. We note again that we use the bar symbol $\overbar{a}$ to stress the fact that the parameters estimated in the learning phase are fixed during the discovery phase. Since the discovery phase relies only on the test data and these parameters, the training set can be discarded.

\subsection{Discovery phase}
The discovery phase looks for novel classes in the test data, given the parameter estimates from the learning phase. Under the D-AMDA modelling framework, we also need to take into account the extra dimensions of the test data. Subsequently, in this phase we need to estimate two main collections of parameters: the parameters of the additional variables corresponding to novel and known classes, and the parameters of the already observed variables related to new and known classes. These characterize the distribution in~\eqref{eq:joint_test} and are estimated keeping the parameter estimates from the learning phase fixed. 

Because the labels of the test data are unobserved, in this stage we aim to optimize the following log-likelihood: 
\begin{equation}\label{eq:discovery_likelihood}
L( \Y ; \bm{\Theta}_y ) = \sumN \text{log} \left\lbrace \sum_{k=1}^K \tau_k \, \phi ( \y_i \,; \MU^*_k, \SIG^*_k ) + \sum_{h=K+1}^C \tau_h \, \phi ( \y_i \,; \MU^*_h, \SIG^*_h ) \right\rbrace,
\end{equation}
with $\sum_{k=1}^K \tau_k + \sum_{h=1}^H \tau_h = 1$. Crucially, for $k = 1,\,\dots,\,K$, mean and covariance parameters of the $K$ observed classes are partitioned into parameters fixed from the learning phase corresponding to the variables observed in $\X$ and parameters corresponding to the additional variables present in the test data:
\begin{equation}\label{eq:partition}
\MU^*_k = ( \overbar{\MU}_k ~~ \MU^{\text{\tiny Q}}_k )^{'} \qquad \SIG^*_k = \begin{bmatrix}
							\overbar{\SIG}_k & \CC_k\\
							\CC^{'}_k & \SIG^{\text{\tiny Q}}_k
							\end{bmatrix},
\end{equation}
where $\CC_k$ are the covariance terms between additional and observed variables. Such partition of the parameters of the observed classes will need to be taken into account during the estimation procedure, as it indirectly induces a constraint on the estimation of the parameters for the additional variables; see the following Section \ref{mstep_ice}.

Optimization of this log-likelihood in the discovery phase is carried out by resorting to an EM algorithm \cite{dempster:etal:1977,mclachlan:krishnan:2008}. From \eqref{eq:discovery_likelihood} we have the complete log-likelihood:
\begin{equation}\label{eq:discovery_complikelihood}
  L ( \Y, \bm{\ell} ; \bm{\Theta}_y ) = \sumN \left[ \sum_{k=1}^K \ell_{ik}\, \text{log} \Big\lbrace \tau_k \, \phi ( \y_i \,; \MU^*_k, \SIG^*_k ) \Big\rbrace + \sum_{h=K+1}^C \ell_{ih}\, \text{log} \Big\lbrace \tau_h \, \phi ( \y_i \,; \MU^*_h, \SIG^*_h  ) \Big\rbrace \right],
\end{equation}
where, $\ell_{ik}$ and $\ell_{ih}$ denote the latent class membership indicators to be estimated on the test data for known and unobserved classes. The EM algorithm alternates the following two steps.

\begin{itemize}
 \item \textbf{E Step}: After estimation of the parameters at the previous M step iteration, the estimated conditional probabilities, $t_{ic} = \widehat{\Pr} \left( \ell_{ic} = 1 \vbar \y_i \right)$ are computed as:
 $$
 t_{ic} = \dfrac{ \widehat{\tau}_c\,\phi(\y_i \,; \widehat{\MU}^*_k, \widehat{\SIG}^*_k) }%
 { \sum_{k=1}^K \widehat{\tau}_k\,\phi(\y_i \,; \widehat{\MU}^*_k, \widehat{\SIG}^*_k) + \sum_{h=K+1}^C \widehat{\tau}_h\,\phi(\y_i \,; \widehat{\MU}^*_h, \widehat{\SIG}^*_h) },
 $$
 for $i=1,\,\dots,\,N$, and $c = 1,\,\dots,\,C$.
 \item \textbf{M Step}: in this step of the algorithm we maximize the expectation of the complete log-likelihood computed using the estimated probabilities $t_{ic}$ of the E step. Due to the augmented dimensions of the test data, this step is more involved. The optimization procedure of the M-step is divided into two parts: estimation of mixing proportions and mean and covariance parameters corresponding to the unobserved classes, described in Section \ref{mstep_mixing}, and estimation of mean and covariance parameters related to the classes already observed in the learning phase, described in Section \ref{mstep_ice}. 
\end{itemize}

\subsubsection{Estimation of mixing proportions and parameters of unobserved classes}
\label{mstep_mixing}
The introduction of new variables does not affect the estimation of the mixing proportions, nor the estimation of the parameters corresponding to the new classes. Hence, in this case the updates are in line with those outlined in \cite{bouveyron:2014}. From equation \eqref{eq:discovery_complikelihood}, the estimates of mean and covariance parameters of the $H$ hidden classes are obtained simply by optimizing the term involving $\MU_h^*$ and $\SIG_h^*$. Therefore, the estimates of the Gaussian density parameters related to the unknown classes are simply given by:
 $$
  \widehat{\MU}^*_h = \dfrac{1}{N_h} \sumN t_{ih} \y_{i},\qquad 
  \widehat{\SIG}^*_h = \dfrac{1}{N_h} \sumN t_{ih} (\y_i - \widehat{\MU}^*_h)(\y_i - \widehat{\MU}^*_h)^{'},
 $$
with $N_h = \sum_i t_{ih}$. For the mixing proportions, two alternative updates are available. One is based on the re-normalization of the mixing proportions $\overbar{\tau}_k$ as outlined in \cite{bouveyron:2014}, the other on the re-estimation of the mixing proportions for both observed and unobserved classes on the test data. The two updates correspond to very different assumptions about the data: the re-normalization update is based on the assumption that the new classes do not affect the balance of the classes observed in the training set, while the other update is based on the assumption that the class proportions may have changed in the test data. We opt for this latter approach, since it is more flexible and avoids the introduction of possible bias due to the re-normalization, as discussed in \cite{lawoko:1989}. The mixing proportions are updated as follows:
$$
\widehat{\tau}_k = \dfrac{N_k}{N} \qquad \widehat{\tau}_h = \dfrac{N_h}{N} \quad \text{for}\quad k = 1,\,\dots,\,K \quad \text{and} \quad h = K+1,\,\dots,\,C.
$$

\subsubsection{Inductive conditional estimation procedure}
\label{mstep_ice}
The estimation of mean and covariance parameters $\MU_k^*$ and $\SIG_k^*$ of the classes already observed in the training data is an involved problem, due to the augmented parameter dimensions and the fact that the parameters from the learning phase needs to be kept fixed. Here we need to estimate the components $\MU^{\text{\tiny Q}}_k$, $\SIG^{\text{\tiny Q}}_k$ and $\CC_k$ of the partitions in \eqref{eq:partition}. As in a standard Gaussian mixture model, a straightforward updated for these would be computing the related in sample weighted quantities. However, this would not take into account the constraint that the parameters $\overbar{\MU}_k$ and $\overbar{\SIG}_k$ have already been estimated in the learning phase and need to be held fixed. In particular,
the covariance block $\overbar{\SIG}_k$ has been estimated in the learning phase via the EDDA model, imposing constraints on its eigen-decomposition. As it is often the case, if the covariance model for $\overbar{\SIG}_k$ has a particular structure (i.e. is not the VVV using the \texttt{mclust} nomenclature) the approach would not ensure a valid positive definite $\SIG_k^*$. A clear example is the case where the EDDA model estimated in the learning phase is a spherical one with diagonal matrices $\overbar{\SIG}_k$. In such case, completing the off-diagonal entries of $\SIG^*_k$ with non-zero terms and without taking into account the structure of the block $\overbar{\SIG}_k$ would not guarantee a positive definite covariance matrix \citep{zhang:2006}. We propose the following procedure to obtain valid estimates.

Denote an observation of the test data $\y_i = \{\y_i^{\text{\tiny P}}, \y_i^{\text{\tiny Q}}\}$, where $\y_i^{\text{\tiny P}}$ are the measurements of the set ${\cal P}$ of variables of the training data and $\y_i^{\text{\tiny Q}}$ are the measurements of the set ${\cal Q}$ of additional variables observed in the test data. To take into account the structure of the block $\overbar{\SIG}_k$, the problem of maximizing the expectation of \eqref{eq:discovery_complikelihood}  with respect to $\SIG^*_k$ and $\MU^*_k$ can be interpreted as the problem of finding estimates of $\MU^{\text{\tiny Q}}_k, \SIG^{\text{\tiny Q}}_k,$ and  $\CC_k$ such that the joint distribution of observed and extra variables $(\{\y_i^{\text{\tiny P}}, \y_i^{\text{\tiny Q}}\} \vbar l_{ik} = 1) \sim {\cal N}(\MU_k^*, \SIG_k^*)$ is a multivariate Gaussian density whose marginal distributions are $(\y_i^{\text{\tiny P}} \vbar l_{ik} = 1) \sim \mathcal{N} (\overbar{\MU}_k, \overbar{\SIG}_k)$ and $(\y_i^{\text{\tiny Q}} \vbar l_{ik} = 1) \sim \mathcal{N} (\MU^{\text{\tiny Q}}_k, \SIG^{\text{\tiny Q}}_k)$, and with $\SIG^*_k$ being positive definite. To accomplish this task, we devise the following \emph{inductive conditional estimation} procedure:
\begin{description}[noitemsep]
 \item[\normalfont\emph{Step 1.}] Fix the marginal distribution of the variables observed in the learning phase, $(\y_i^{\text{\tiny P}} \vbar l_{ik} = 1) \sim \mathcal{N} (\overbar{\MU}_k, \overbar{\SIG}_k)$;
 \item[\normalfont\emph{Step 2.}] Estimate the parameters of the conditional distribution\, $(\y_i^{\text{\tiny Q}} \vbar \y_i^{\text{\tiny P}}, l_{ik} = 1) \sim {\cal N}(\mathbf{m}_{ik}, \mathbf{E}_k)$, where $\mathbf{m}_{ik}$ and $\mathbf{E}_k$ are related mean and covariance parameters.
 \item[\normalfont\emph{Step 3.}] Find estimates of the parameters of the joint distribution $(\{\y_i^{\text{\tiny P}}, \y_i^{\text{\tiny Q}}\} \vbar l_{ik} = 1) \sim {\cal N}(\MU_k^*, \SIG_k^*)$ using the fixed marginal and the conditional distribution. 
\end{description}
Since we are using an inductive approach, \emph{Step 1} corresponds in keeping $\overbar{\MU}_k, \overbar{\SIG}_k$ fixed. Next, in \emph{Step 2} the parameter estimates of the distribution of the new variables given the variables observed in the training set are obtained. This allows to take into account the information and the structure of the learning phase parameters. Then, in \emph{Step 3} these estimates are used to find the parameters of the marginal distribution of the set of new variables ${\cal Q}$ and the joint distribution of ${\cal R} = \{ {\cal P}, {\cal Q}\} $, while preserving the joint association structure among all the variables in ${\cal R}$. The proposed method is related to the well known \emph{iterative proportional fitting} algorithm for fitting distributions with fixed marginals \citep[see for example][]{whittaker:1990,fienberg:meyer:2006}, and the \emph{iterative conditional fitting} algorithm of \cite{chaudhuri:etal:2007} used to estimate a multivariate Gaussian distribution with association constraints.

Taking the expectation of the complete log-likelihood in \eqref{eq:discovery_complikelihood}, the term involving $\MU^*_k$ and $\SIG^*_k$ can be rewritten as:
\begin{equation}\label{eq:conditional}
 \sumN \left[ \, \sum_{k=1}^K t_{ik}\, \text{log} \Big\lbrace \, \phi ( \yq_i \vbar \yp_i \,;\, \mathbf{m}_{ik}, \mathbf{E}_k  ) \, \phi ( \yp_i \,;\, \overbar{\MU}_k, \overbar{\SIG}_k ) \Big\rbrace \right].
\end{equation}
In \emph{Step 1}, parameters $\overbar{\MU}_k, \overbar{\SIG}_k$ are fixed from the learning phase. Therefore, the term $\text{log}\lbrace\phi ( \yp_i \,;\, \overbar{\MU}_k, \overbar{\SIG}_k)\rbrace$ is already maximized. In \emph{Step 2} and \emph{Step 3}, we make use of the well known closure properties of the multivariate Gaussian distribution \citep[see][for example]{tong:1990,zhang:2006} in order to maximize the term $\log \lbrace \phi ( \yq_i \vbar \yp_i \,;\, \mathbf{m}_k, \mathbf{E}_k  ) \rbrace$. In {\em Step 2} the focus is on the conditional distribution; for each observation $i$ we can rewrite:
$$
\mathbf{m}_{ik} = \MU^{\text{\tiny Q}}_k + \CC_k^{'} \overbar{\SIG}^{-1}_k (\yp_i - \overbar{\MU}_k ), \qquad 
\mathbf{E}_k = \SIG^{\text{\tiny Q}}_k - \CC_k^{'} \, \overbar{\SIG}^{-1}_k \, \CC_k.
$$
Let us define the scattering matrix $\OO_k = \sumN t_{ik} ( \y_i - \overbar{\y}_k )(\y_i - \overbar{\y}_k)^{'}$, with $\overbar{\y}_k = \frac{1}{N_k} \sumN t_{ik}\, \y_i$. We can partition it as:
$$
\OO_k = \begin{bmatrix}
                   \W_k & \mathbf{V}_k\\
                   \mathbf{V}^{'}_k & \mathbf{U}_k
               \end{bmatrix},
$$
with $\W_k$ the block related to the variables observed in the learning set, $\mathbf{U}_k$ the block associated to the new variables and $\mathbf{V}_k$ the crossproducts. Now we maximize~\eqref{eq:conditional} with respect to $\mathbf{E}_k$ and $\CC_k$. After some algebraic manipulations, we obtain the estimates:
$$
\widehat{\CC}_k = ( \overbar{\SIG}^{-1}_k \, \W_k \, \overbar{\SIG}^{-1}_k )^{-1} ( \overbar{\SIG}^{-1}_k \mathbf{V}_k)
$$
$$
\widehat{\mathbf{E}}_k = \dfrac{1}{N_k} \left[ \widehat{\CC}^{'}_k \, \overbar{\SIG}^{-1}_k \, \W_k \, \overbar{\SIG}^{-1}_k \, \widehat{\CC}_k - 
2 \mathbf{V}^{'}_k \overbar{\SIG}^{-1}_k \, \widehat{\CC}_k + \mathbf{U}_k \right],
$$
Then, in \emph{Step 3} we obtain the estimates of the marginal distribution for the set of extra variables as:
$$
\widehat{\MU}^{\text{\tiny Q}}_k = \dfrac{1}{N_k} \left[\, \sumN t_{ik} \yq_i - \widehat{\CC}^{'}_k \, \overbar{\SIG}^{-1}_k \sumN t_{ik} (\yp_i - \overbar{\MU}_k ) \right], \qquad
\widehat{\SIG}^{\text{\tiny Q}}_k = \widehat{\mathbf{E}}_k + \widehat{\CC}_k^{'} \, \overbar{\SIG}^{-1}_k \, \widehat{\CC}_k.
$$
Hence $\MU^*_k$ and $\SIG^*_K$ are estimated:
$$
\widehat{\MU}^{*}_k = ( \overbar{\MU}_k ~~ \widehat{\MU}^{\text{\tiny Q}}_k )^{'}, \qquad
\widehat{\SIG}^{*}_k = \begin{bmatrix}
			      \overbar{\SIG}_k & \widehat{\CC}_k\\
			      \widehat{\CC}^{'}_k & \widehat{\SIG}^{\text{\tiny Q}}_k
			   \end{bmatrix}.
$$
Further details about the derivations are in Appendix~\ref{app:est}. Provided that $\OO_k$ is positive definite, the estimate of $\SIG_k$ obtained in such way is ensured to be positive definite as well due to the properties of the Schur complement \citep{zhang:2006,tong:1990}. In certain cases, for example when the number of variables $R$ is large compared to $N$ and to the expected class sizes, or when the variables are highly correlated, this scattering matrix could be singular. To overcome this issue, one could resort to regularization. To this purpose, we delineate a simple Bayesian regularization approach in Appendix~\ref{app:regularization}.

\subsection{Initialization and selection of the number of hidden classes}
In order to compute the first E step iteration of the EM algorithm in the discovery phase, we need to initialize the parameter values. A random initialization has a fair chance of not providing good starting points. On the other hand, the initialization based on the model-based hierarchical clustering method discussed in \citep{scrucca:raftery:2015} and \citep{fraley:1998} often yields good starting points, is computationally efficient and works well in practice. However, we need to take care of the fact that a subset of the parameters is fixed. We make use of the following strategy for initialization. 

First we obtain a hierarchical unsupervised partition of the observations in the test data using the method of \citep{scrucca:raftery:2015} and \citep{fraley:1998}. Afterwards, for a fixed number $C$ of clusters and the corresponding partition, we compute the within-cluster means and covariance matrices, both for new and observed variables. Let us denote with $\tilde{\MU}^{\text{\tiny P}}_g$ and $\tilde{\SIG}^{\text{\tiny P}}_g$ $(g=1,\dots,C)$ the computed cluster parameters related to the observed variables, with $\tilde{\MU}^{\text{\tiny Q}}_g$ and $\tilde{\SIG}^{\text{\tiny Q}}_g$ those related to the extra variables, and with $\tilde{\CC}_g$ the covariance terms. Now, we find which of the detected clusters match the classes observed in the training set over the observed variables. For each known class and each cluster we compute the Kullback-Leibler divergence:
$$
\text{tr} \left\{ \left(\tilde{\SIG}^{\text{\tiny P}}_g\right)^{-1} \overbar{\SIG}_k \right\} + ( \tilde{\MU}^{\text{\tiny P}}_g - \overbar{\MU}_k )^{'} \left(\tilde{\SIG}^{\text{\tiny P}}_g\right)^{-1} ( \tilde{\MU}^{\text{\tiny P}}_g - \overbar{\MU}_k )+ \log \dfrac{\det\tilde{\SIG}^{\text{\tiny P}}_g}{\det\overbar{\SIG}_k}, \qquad \forall ~g, k.
$$
Then, we find the first $K$ clusters with the minimum divergence and thus likely corresponding to the classes observed in the training data. For these clusters, the set of parameters related to the observed variables are initialized with the associated values $\overbar{\MU}_k$ and $\overbar{\SIG}_k$, the set of parameters related to the new variables are initialized with the same values $\tilde{\MU}^{\text{\tiny Q}}_k$ and $\tilde{\SIG}^{\text{\tiny Q}}_k$, and the covariance terms with $\tilde{\CC}_k$. The remaining clusters can be considered as hidden classes and the related parameters are initialized with the corresponding cluster means and covariances.

Similarly to AMDA, also in the D-AMDA framework class detection corresponds corresponds to selection of the number of hidden classes in the test data. As in the learning phase, the BIC is employed for this purpose. Explicitly, for a range of values of number of hidden classes $H$, we choose the model that maximizes the quantity:
$$
\text{BIC}_H = 2\, L ( \Y ; \widehat{\bm{\Theta}}_y ) - \eta_H \log N,
$$
where $\eta_H$ is the number of parameters estimated in the discovery phase, equal to $(H + K - 1) + 2HR + H \binom{R}{2} + 2KQ + KPQ + K\binom{Q}{2}$.

\section{Inductive variable selection for D-AMDA}
\label{varsel}

Given the large amount and the variety of sources at disposition, classification of high-dimensional data is becoming more and more a routine task. In this setting, variable selection has been proven beneficial for increasing accuracy, reducing the number of parameters and a better model interpretation \citep{guyon:2003,pacheco:2006,brusco:2011}. We adapt the variable selection method of \cite{maugis:2011} and \cite{murphy:2010} in order to perform \emph{inductive} variable selection within the context of D-AMDA. 
The aim is to select the relevant variables that contain the most useful information about both observed and novel classes. The method is inductive in the sense that the classifier model first is built on the data observed in the learning phase. Then, while performing variable selection on the new test data, the classifier is adapted by removing and adding variables without re-estimating the model on the learning data.

Following \cite{maugis:2011} and \cite{murphy:2010}, at each step of the variable selection procedure we consider the partition $\Y = (\Y^{\text{class}}, Y^{\text{prop}}, \Y^{\text{other}})$, where $\Y^{\text{class}}$ is the current set of relevant variables, $Y^{\text{prop}}$ is the variable proposed to be added/removed to/from $\Y^{\text{class}}$, and $\Y^{\text{other}}$ are the non relevant variables. Let also $\bm{\ell}$ be the class indicator variable. For each stage of the algorithm, we compare two models:
\[
\begin{split}
  \mathcal{M}_{1} ~\colon ~ &p(\Y\vbar\bm{\ell}) = p( \Y^{\text{class}}, Y^{\text{prop}} \vbar \bm{\ell} ) \, p( \Y^{\text{other}} ), \\
  \mathcal{M}_{2} ~\colon ~ &p(\Y\vbar\bm{\ell}) =  p( \Y^{\text{class}} \vbar \bm{\ell} ) \, p( Y^{\text{prop}} \vbar \Y^{\text{reg}} \subseteq \Y^{\text{class}} ) \, p( \Y^{\text{other}} ). 
\end{split}
\]
In model $\mathcal{M}_{1}$, $\Y^{\text{p}}$ is relevant for classification and $p( \Y^{\text{class}}, Y^{\text{prop}} \vbar \bm{\ell} )$ is the D-AMDA model where the classifier is adapted by including the proposed variable $Y^{\text{prop}}$. In model $\mathcal{M}_{2}$, $Y^{\text{prop}}$ does not depend on the labels and thus is not useful for classification. $p( \Y^{\text{class}} \vbar \bm{\ell} )$ is the D-AMDA model on the current selected variables and the conditional distribution $p( Y^{\text{prop}} \vbar \Y^{\text{reg}} \subseteq \Y^{\text{class}} )$ is a regression where  $Y^{\text{prop}}$ depends on $\Y^{\text{class}}$ through a subset of predictors $\Y^{\text{reg}}$. This regression term encompasses the fact that some variables may be redundant given the set of already selected ones, and thus can be discarded \citep{murphy:2010,raftery:dean:2006}. Relevant predictors are chosen via a standard stepwise procedure and the selection avoids to include unnecessary parameters that would over-penalize the model without a significant increase in its likelihood \citep{maugis:celeux:2009:a,maugis:celeux:2009:b}. The two models are compared by computing the difference between their BIC:
\[
\begin{split}
  \text{BIC}_{1}&= \text{BIC}_{\text{class}}( \Y^{\text{class}}, Y^{\text{prop}} ), \\
  \text{BIC}_{2}&= \text{BIC}_{\text{no class}}( \Y^{\text{class}} ) + \text{BIC}_{\text{reg}}( Y^{\text{prop}} \vbar \Y^{\text{reg}} \subseteq \Y^{\text{class}} ), 
\end{split}
\]
where $\text{BIC}_{\text{class}}( \Y^{\text{class}}, Y^{\text{p}} )$ is the BIC of the D-AMDA model where $Y^{\text{prop}}$ is useful for classification, $\text{BIC}_{\text{no class}}( \Y^{\text{class}} )$ is the BIC on the current set of selected variables and $\text{BIC}_{\text{reg}}( Y^{\text{prop}} \vbar \Y^{\text{reg}}$ $\subseteq \Y^{\text{class}} )$ is the BIC of the regression. The difference $ (\text{BIC}_{1} - \text{BIC}_{2}) $ is computed and if it is greater than zero, there is evidence that $Y^{\text{p}}$ conveys useful information about the classes, hence variable $Y^{\text{prop}}$ is added to the D-AMDA model and the classifier is updated. 

The selection is performed using a stepwise greedy forward search where variables are added and removed in turn. Since we adopt an inductive approach, when the variables to be added/removed belong to the set of variables already observed in the learning phase, the classifier is updated in a fast and efficient way. Indeed, if a variable observed in $\X$ needs to be added, the classifier is updated by simply augmenting the set of parameters with the parameters already estimated in the learning phase. Analogously, if the variable needs to be removed, the classifier is updated by deleting the corresponding parameters. Only parameters related to additional variables and novel classes need to be estimated when updating the D-AMDA model. Parameters related to known classes and observed variables are updated only via deletion or addition. As such, the method is suitable for fast on-line variable selection.

The classification procedure is partly unsupervised because of the presence of unobserved classes. Therefore, while searching for the relevant variables, also the number $H$ of unknown classes needs to be chosen. As in \cite{maugis:celeux:2009:a,maugis:celeux:2009:b,raftery:dean:2006}, we consider a range of possible values for $H$. Then, at every step $\text{BIC}_{\text{class}}( \Y^{\text{class}}, Y^{\text{prop}} )$ and $\text{BIC}_{\text{no class}}( \Y^{\text{class}} )$ are computed by maximizing over this range. 
Therefore, the method returns both the set of relevant variables and the optimal number of unobserved classes. 

The set of relevant variables needs to be initialized at the first stage of the variable selection algorithm. We suggest to start the search from a conveniently chosen subset of size $S$ of the variables observed in the learning phase. To determine such subset, for every variable in $\Y$ corresponding to those already observed in $\X$, we estimate a univariate Gaussian mixture model for a number of components ranging from 0 to $G > K$. Then we compute the difference between the BIC of such model and the BIC of a single univariate Gaussian distribution. The variables are ranked according to this difference from the largest to the lowest value. The starting subset is formed by selecting the top $S$ variables in the list. Similar initial selection strategies have been discussed in \cite{mclachlan:2004} and \cite{murphy:2010}. Note that one could also initialize the set of relevant variables from all the observed variables of the training data. Nonetheless, if the number of variables observed in $\X$ is large, it is likely that many of them would be uninformative or redundant, therefore, initialization using such set might not provide a good starting point for the search.

\section{Simulated data experiments}
\label{sim}
In this section we evaluate the proposed modeling framework for variable selection and adaptive classification through different simulated data experiments under various conditions. The objective is to assess the classification performance of the method, its ability of detecting the novel classes and its ability of discarding irrelevant variables and selecting those useful for classification.

\subsection{Simulation study 1}
\label{sim_1}
This simulation study shows the usefulness of using all the variables available in the test data for class prediction and detection when only a small subset of these are observed in the training stage.

We consider the well known Italian wines dataset \citep{forina:1986}. The data consist of 27 chemical measurements from a collection of wine samples from Piedmont region, in Italy. The observations are classified into three classes indicating the type of wine. Different scenarios are considered for different combinations of number of variables observed in the training stage and different test data sample sizes. Using the class-specific sample means and covariances, we generate training data sets with random subsets of the 27 variables, with the number of variables observed in the training set equal to 18, 9, and 3. Then, with the same class-specific parameters, a test set on all the 27 variables and different sample sizes is generated. One class is randomly deleted from the training data, while all 3 classes are present in the test data. In each scenario we consider the following models: the EDDA classifier fitted on the training data with full information, i.e. all 3 classes and all 27 variables, tested on the full test data; the EDDA classifier fitted on the training data considering only a subset of the variables, then tested on the test data containing the same subset of training variables; the AMDA approach of \cite{bouveyron:2014} fitted on the simulated training data with a subset of the variables and tested on the test data with the subset of variables observed in the training; the presented D-AMDA framework.

Details are described Appendix \ref{app:sim}, with detailed results reported in Figures \ref{fig_sim_wine_P1_err}, \ref{fig_sim_wine_P1_ari}, \ref{fig_sim_wine_P2_err}, \ref{fig_sim_wine_P2_ari}, \ref{fig_sim_wine_P3_err}, and \ref{fig_sim_wine_P3_ari}. 
The variables in the wine data present a good degree of discrimination, and the EDDA model fitted and tested on the complete data represents the optimal baseline performance. On the other hand, the EDDA classifier trained on the partial data cannot account for the unobserved class in the test data and provides the worst classification performance.
AMDA can detect additional classes in the test data, but it cannot use the discriminant information potentially available in the additional variables, thus obtaining an inferior classification performance compared to D-AMDA. Since the D-AMDA framework adapts to the additional dimensions and classes, all the information available in the variables observed in the test set is exploited for classification, of both variables observed during the training stage and the extra ones present in the test set. This extra information is beneficial, especially when the number of variables present in the training set is small ($P = 3$ in particular), attaining a classification performance comparable to the optimal baseline.

\subsection{Simulation study 2}
\label{sim_2}
This simulation study assesses the D-AMDA classification performance and the effectiveness of the inductive variable selection method at detecting variables relevant for classification. Different scenarios are constructed by defining different proportions of relevant and irrelevant variables available in the training and the full test data.

In all the experiments of this section we consider three types of variables: class-generative variables (\Gen), which contain the principal information about the classes, redundant variables (\Cor), which are correlated to the generative ones, and noise variables (\Noi), which do not convey any information about the classes. The \Gen{}  variables are distributed according to mixture of $C=4$ multivariate Gaussian distributions. Each \Cor{} variable is correlated to 2 \Gen{} variables selected at random, while the \Noi{} variables are independent from both \Gen{} and \Cor{} variables. In the learning set, 2 of the 4 classes are observed and they are randomly chosen. All the 4 classes are observed in the test set. Further details about the parameters of the simulations are in Appendix~\ref{app:sim}. Three experiments are considered, each one characterized by three scenarios.

We point out the fact that, as they are generated, the \Cor{} variables actually contain some information about the classification. Indeed, they are independent of the label variable only conditionally on the set \Gen{}, not marginally. Thus, in some cases, they could convey the best information available to classify the data units if some generative variables have been discarded during the search. Hence, the inclusion of a \Cor{} variable would not necessarily degenerate the classification performance.

\subsubsection{Experiment 1}
The test data consist of 100 variables, 10 \Gen{}, 30 \Cor{} and 60 \Noi{}. In the learning set, 20 of the 100 variables are observed. Three scenarios are defined according to the set of variables observed in the simulated $\X$:
\begin{itemize}[noitemsep]
 \item[(1.a)] All the 10 \Gen{} variables plus 10 variables picked at random among \Cor{} and \Noi{}.
 \item[(1.b)] 5 \Gen{} variables selected at random, 5 \Cor{} selected at random, plus 10 variables chosen at random among \Cor{} and \Noi{}.
 \item[(1.c)] 2 \Gen{} selected at random, the remaining 18 variables are chosen at random among \Cor{} and \Noi{}.
\end{itemize}
The sample size of the learning set is equal to the sample size of $\Y$ and takes values 100, 200 and 400. In all scenarios, the forward search is initialized starting from all the variables observed in $\X$.

In (1.a), the EDDA learning model is estimated on a set containing all the classification variables. Furthermore, the forward search is initialized on the same set. This gives a good starting point to the variable selection procedure, resulting that only \Gen{} variables are declared as relevant and with an excellent classification performance. The results hold regardless of the size of the samples in practice (Appendix~\ref{app:sim}, Figures~\ref{fig_sim1_s1} and \ref{fig_sim1_s1_ari}). As less \Gen{} variables are available in the learning phase, the variable selection method declares as relevant \Cor{} variables more frequently. However, good classification results and good selection performance are still obtained, especially for larger sample sizes (Appendix~\ref{app:sim}, Figures~\ref{fig_sim1_s2}, \ref{fig_sim1_s3}, \ref{fig_sim1_s2_ari}, \ref{fig_sim1_s3_ari}).

\subsubsection{Experiment 2}
Also here the test data consist of 100 variables, 10 \Gen{}, 30 \Cor{} and 60 \Noi{}. In the learning set, 50 of the 100 variables are observed. Three scenarios are defined according to the set of variables observed in the simulated $\X$:
\begin{itemize}[noitemsep]
 \item[(2.a)] All the 10 \Gen{} variables plus 40 variables randomly chosen among \Cor{} and \Noi{}.
 \item[(2.b)] 5 \Gen{} variables selected at random, 15 \Cor{} selected at random, plus 30 variables chosen at random among \Cor{} and \Noi{}.
 \item[(2.c)] 2 \Gen{} selected at random, the remaining 48 variables are randomly selected among \Cor{} and \Noi{}.
\end{itemize}
In this experiment, the sample size of the learning set is fixed and equal to 50 for all the scenarios. The forward search is initialized from 10 of the 50 variables observed in $\X$, selected using the procedure described in Section~\ref{varsel}. 

This setting is particularly challenging, since the learning set is high-dimensional in comparison with the number of data points. In practice, this results in a learning phase where only EDDA models with diagonal covariance matrices can be estimated. Even if all \Gen{} variables are observed in $\X$, such subset of models are misspecified in relation to how the data is generated. This represents a difficult starting point for the D-AMDA model and the variable selection procedure. Indeed, with this experiment we want to test the robustness of the method against the misspecification of the model in the learning stage. Nevertheless, in scenario (2.a) a selection of reasonable quality is attained. In all three scenarios, \Noi{} variables are never selected and the method achieves a good classification performance even if \Cor{} variables are selected as relevant almost as many times as the variables of the \Gen{} set (Appendix~\ref{app:sim}, Figures~\ref{fig_sim2_s1}, \ref{fig_sim2_s2}, \ref{fig_sim2_s3}, \ref{fig_sim2_s1_ari}, \ref{fig_sim2_s2_ari}, \ref{fig_sim2_s1_ari}). This fact is likely due to the variable selection initialization: this initialization strategy tends to start the selection from a set of good classification variables, and such set may contain both \Gen{} and \Cor{} variables.

\subsubsection{Experiment 3}
In this case the test data consist of 200 variables, 20 \Gen{}, 60 \Cor{} and 120 \Noi{}. In the learning set, 40 of the 200 variables are observed. Three scenarios are defined according to the set of variables observed in the simulated $\X$:
\begin{itemize}[noitemsep]
 \item[(3.a)] All the 20 \Gen{} variables plus 20 variables selected randomly among \Cor{} and \Noi{}.
 \item[(3.b)] 10 \Gen{} variables selected at random, 10 \Cor{} selected at random, plus 20 variables picked at random among \Cor{} and \Noi{}.
 \item[(3.c)] 4 \Gen{} selected at random, the remaining 36 variables are randomly chosen among \Cor{} and \Noi{}.
\end{itemize}
Here, the sample size of the learning set is equal to the one of the test data and takes values 100, 200 and 400. The forward search is initialized from 10 of the 40 variables observed in $\X$, selected using the procedure described in Section~\ref{varsel}. 

The experiment is characterized by $\Y$ being high-dimensional. For larger sample sizes, the variable selection method tends to correctly identify the relevant variables, especially as the number of \Gen{} variables involved in the estimation of the EDDA model in the learning phase increases. \Cor{} are declared as relevant more often when such number is reduced. However, \Noi{} variables are never selected (Appendix~\ref{app:sim}, Figures~\ref{fig_sim3_s1}, \ref{fig_sim3_s2}, \ref{fig_sim3_s3}). The D-AMDA with variable selection obtains good classification results in all the scenarios  (Appendix~\ref{app:sim}, Figures~\ref{fig_sim3_s1_ari}, \ref{fig_sim3_s2_ari}, \ref{fig_sim3_s3_ari}).

\section{Contaminated honey data}

\begin{figure}[t]
 \centering
 \includegraphics[scale=0.8]{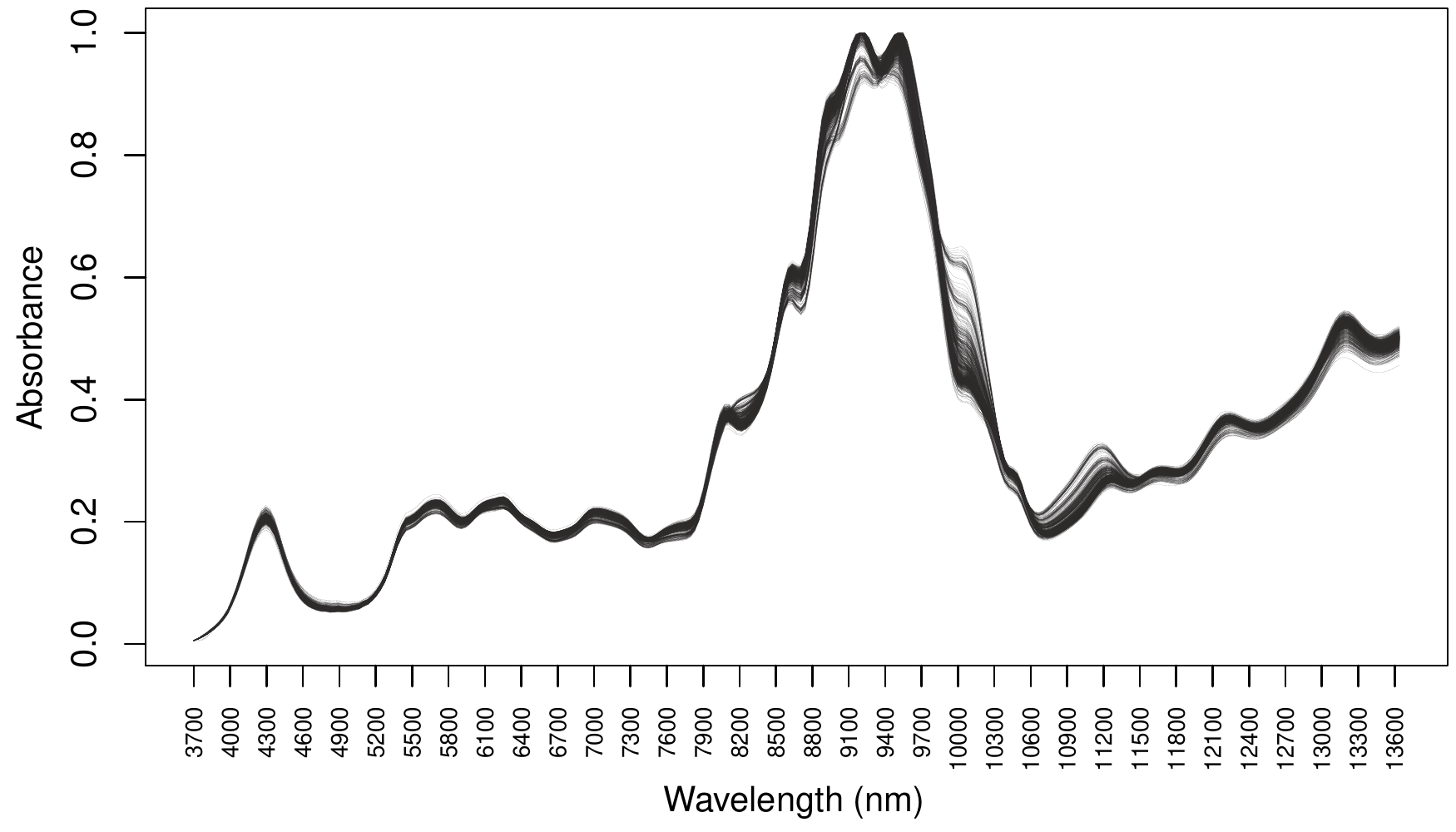}
 \caption{\label{honey1} The Mid-infrared spectra recorded for the pure and the contaminated honey samples.}
\end{figure}

\begin{figure}[!h]
 \centering
 \includegraphics[scale=0.8]{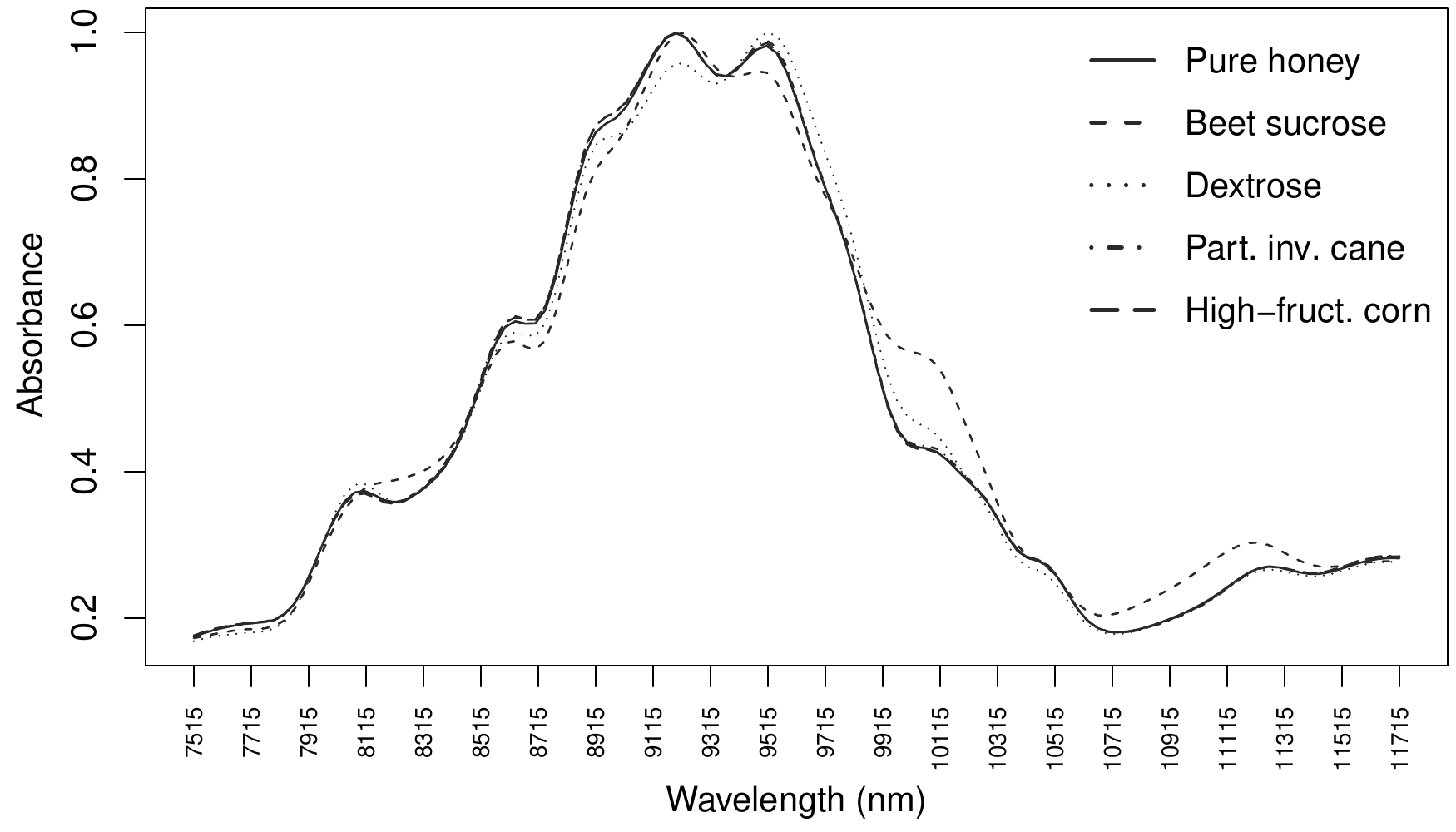}
 \caption{\label{honey2} Class-conditional mean spectra for pure and contaminated honey samples. The figure is a zoom on the range $7500nm-11700nm$ approximatively.}
\end{figure}

Food authenticity studies are concerned with establishing whether foods are authentic or not. Mid-infrared spectroscopy provides an efficient method of collecting data for use in food authenticity studies, without destructing the sample being tested nor requiring complex preparation \citep{downey:1996}. In this section we consider a food authenticity data set consisting of mid-infrared spectroscopic measurements of honey samples. \citep{kelly:2006} collected 1090 absorbance spectra of artisanal Irish honey over the wavelength range $3700nm-13600nm$ at $35nm$ interval. Therefore, the data consists of 285 absorbance values (variables). Of these samples, 290 are pure honey, while the remaining are contaminated with five sugar syrups: beet sucrose (120), dextrose syrup (120), partial invert cane syrup (160), fully inverted beet syrup (280) and high-fructose corn syrup (120). The aim is to discriminate the pure honey from the adulterated samples and the different contaminants. At the same time, the purpose is in the identification of a small subset of absorbance values containing as much information for authentication purposes as the whole spectrum does. Figures~\ref{honey1} and~\ref{honey2} provide a graphical description of the data. Except from beet sucrose and dextrose, there is an high overlapping between the other contaminants and the pure honey; this stems from the similar composition of honey and these syrups \citep{kelly:2006}. The principal features seem to be around the ranges $8700nm-10300nm$ and $10500nm-11600nm$, while the spectra overlap significantly at lower wavelengths.

\label{appl}

In this section we test the D-AMDA method with variable selection. We construct an artificial experiment that represents the situation were the samples in the learning set were collected at a lower resolution than the ones in test data and the information about one of the contaminants was missing. We randomly split the whole data into learning set and test set, in proportions $2/3$ and $1/3$ respectively. Then, we consider the learning set as it were generated from absorbance spectra collected at $70nm$ intervals, retaining wavelengths $3700nm$, $3770nm$, $3840nm$ and so on. Thus, the data observed in the learning phase are approximately recorded on half of the variables of the test data. Afterwards, we randomly chose one of the two classes related to the contaminants beet sucrose and dextrose syrup, and we remove from the learning set the corresponding observations. In this way we obtain a test set measured on additional variables and containing extra classes. 

We replicate the experiment for 100 times, applying the D-AMDA approach with variable selection (\textsf{varSel}) and without (\textsf{noSel}). For comparison and for evaluating the classification performance, we also apply the EDDA model (\textsf{orac}) to the whole learning set containing all the contaminants. Then we use the estimated classifier on the test data to classify the samples. The EDDA classifier uses all the information available about classes and wavelengths, thus its classification performance can be considered as the ``oracle'' baseline. For the variable selection, we initialize the search from a set of 30 wavelengths selected as described in Section~\ref{varsel}.

\begin{figure}[!t]
 \centering
 \includegraphics[scale=0.8]{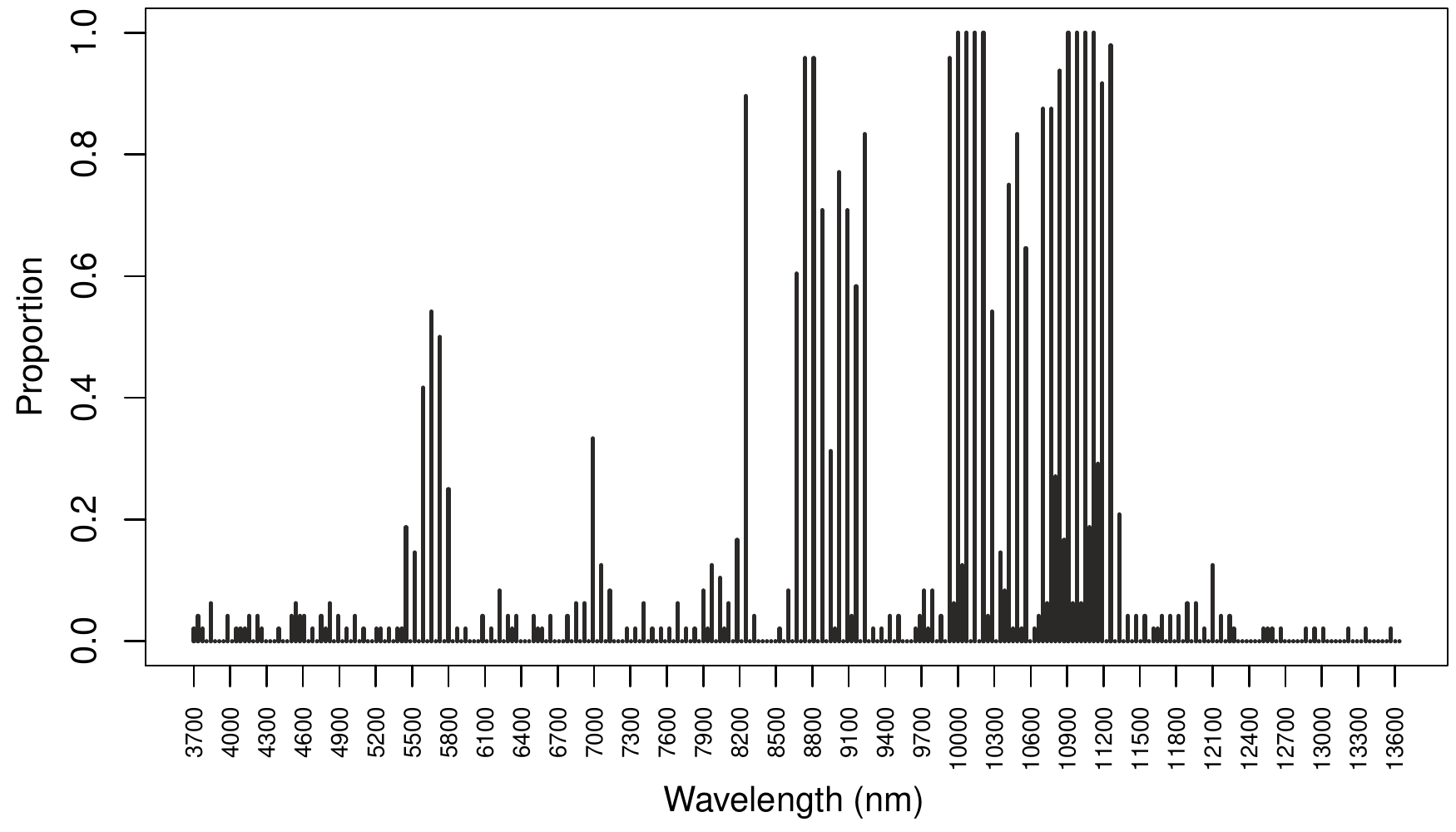}
 \caption{\label{honey3} Proportions of time a wavelength has been selected as a relevant variable over a hundred replicates of the artificial experiment.}
\end{figure}

\begin{figure}[!t]
 \centering
 \subfloat[][]
 {\includegraphics[scale=0.5]{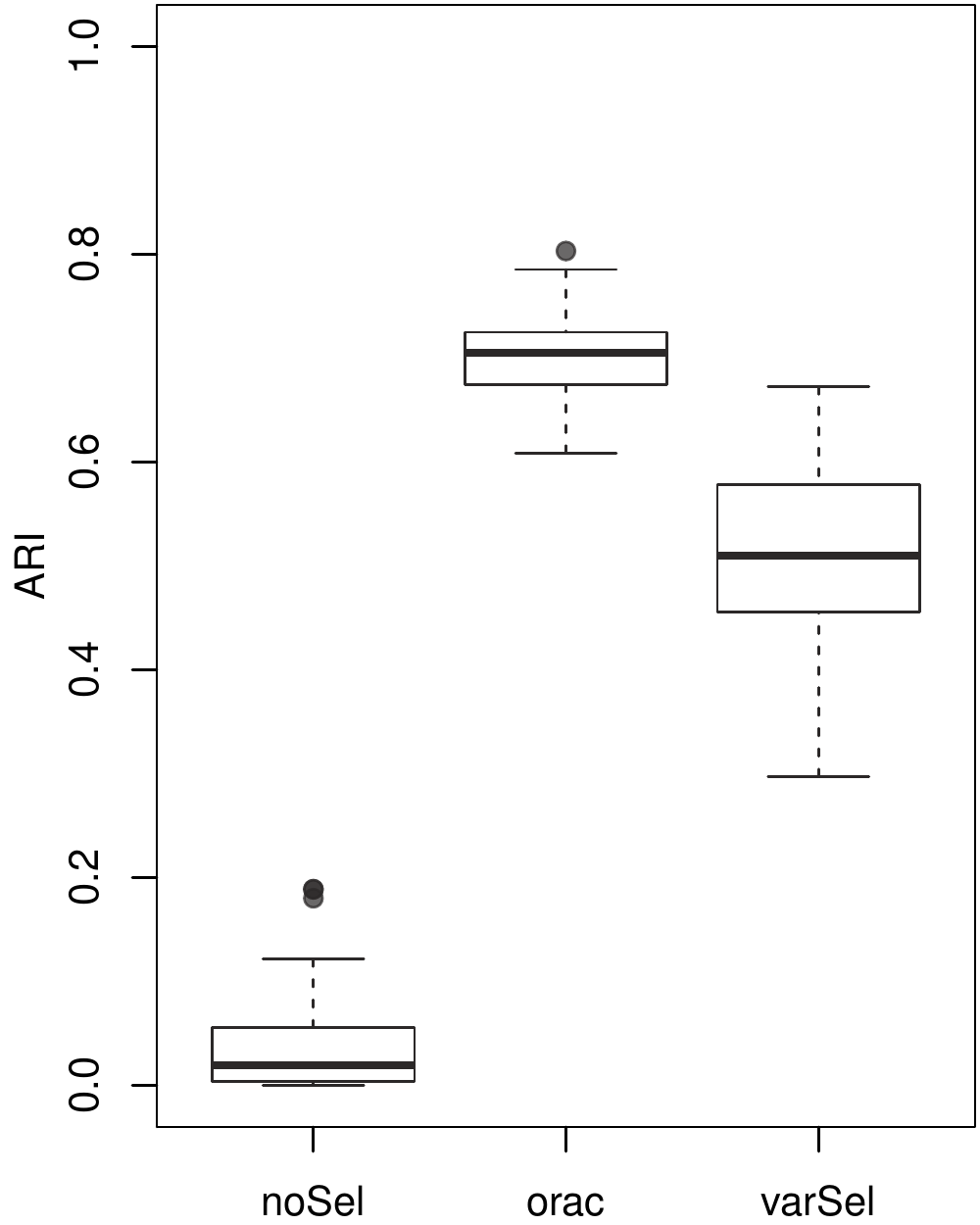}} \qquad 
 \subfloat[][]
 {\includegraphics[scale=0.5]{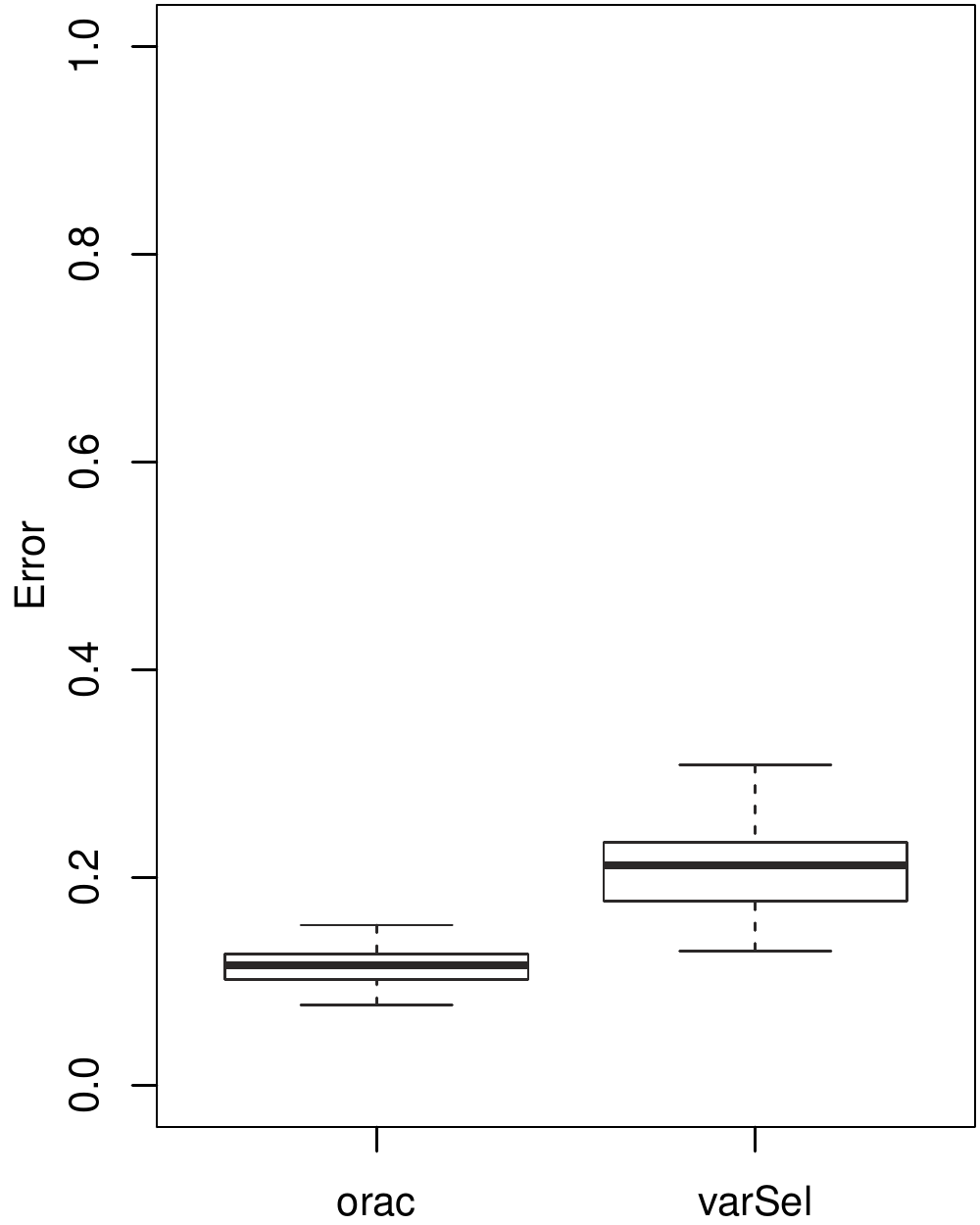}}
 \caption{\label{honey4} ARI (a) and classification error (b). The classification error is reported for the EDDA model and the D-AMDA with variable selection. For the error, the boxplot displays values only for the 79/100 times the D-AMDA correctly selected the number of unknown classes.}
\end{figure}

Results of the variable selection procedure are reported in Figure~\ref{honey3}. The figure displays the proportion of times a wavelength has been declared as relevant for separating the classes of contaminants and the pure honey. The frequently chosen wavelengths are mostly in the ranges $10000nm - 11200nm$ and $8500nm-9300nm$. In particular, values $10000nm$, $10070nm$, $10140nm$, $10210nm$, $10910nm$, $10980nm$, $11050nm$, and $11120nm$ are selected in all the replicates of the experiment. Also wavelengths in the range $5400nm-5800nm$ are selected a significant number of times. The peak in $8250nm$ corresponds to a wavelength range particularly useful to discriminate dextrose syrup from the rest \citep{kelly:2006}. The most frequently selected wavelengths correspond to the interesting peaks and features of the spectra. Classification results are presented in Figure~\ref{honey4}. As for the simulation settings, because of the extra hidden class in the test set we made use of the ARI \citep{hubert:arabie:1985} to compare the actual classification and the ones estimated by \textsf{varSel} and \textsf{noSel} (see also Appendix \ref{app:sim}). \textsf{noSel} selects the correct number of classes only 34/100 of the times. \textsf{varsSel} chose the right number of unknown classes 79 out of 100 times and panel (b) of Figure~\ref{honey4} reports the boxplot of the classification error of \textsf{varSel} and \textsf{orac} in this case. The classification performance of \textsf{varsel} is comparable to \textsf{orac}, but it makes use of information about less wavelengths and is obtained in a more complex setting.

\section{Discussion}
\label{disc}

We presented a general adaptive mixture discriminant analysis method for classification and variable selection when the test data contain unobserved classes and extra variables. We have shown that our methodology effectively addresses the issues generated by the presence of hidden classes in a test data with augmented dimensions compared to the data observed during the training stage. As such, the method is suitable for applications in real-time classification problems where the new data points to be labelled convey extra information thanks to the presence of additional input features.

The inductive approach had the advantage of avoiding the storing of the learning set and of avoiding the re-estimation of the parameters already obtained in the learning stage. However, when extra variables are observed in the test data, the estimation process is a complex problem, due to the parameter constraints induced by the initial learning phase. An inductive conditional estimation procedure has been introduced to overcome the issue and obtain valid parameter estimates related to the added dimensions. 
The inductive framework results in a fast and computationally efficient procedure, which has been embedded into a variable selection method for dealing with high-dimensional data. 

The method proposed in this paper opens interesting future research directions. A limitation of the D-AMDA framework is that the discovery phase does not consider particular constraints on the estimated covariance matrices, reducing the flexibility of the model. The introduction of parsimonious models as in \cite{bensmail:celeux:1996} with adaptive dimensions is object of current investigation. Another limitation is that the labels observed in the training data are assumed to be noise free, as well as that no outlier observations are present in the input features. Recent work by \cite{cappozzo:2020} proposes a robust version of the AMDA framework to address these added sources of complexity. Future work may explore the development of a robust version of D-AMDA, with a particular focus on discarding those additional dimensions characterized by high levels of noisy and contaminated observations, suitable for robust on-line classification.

\newpage
\section*{Appendix}
\appendix

\section{Details of the inductive conditional estimation}
\label{app:est}
Neglecting the term involving the mixing proportions, the objective function to be optimized in the M step to estimate $\MU_k^*$ and $\SIG^*_k$ is given by: 
$$
F(\MU^*, \SIG^*) = \sumN \left[\, \sum_{k=1}^K t_{ik}\, \text{log} \Big\lbrace \phi ( \y_i \,; \MU^*_k, \SIG^*_k ) \Big\rbrace \right].
$$
Let $\OO_k = \sumN t_{ik} ( \y_i - \overbar{\y}_k )(\y_i - \overbar{\y}_k)^{'}$, with $\overbar{\y}_k = \frac{1}{N_k} \sumN t_{ik}\, \y_i$. The above function can be expressed in term of the covariance matrix as:
$$
F(\SIG^*) = \sum_k \text{tr} \bigl\{ \OO_k (\SIG_k^*)^{-1}\bigr\} +\sum_k N_k \log \det \SIG_k^*.
$$
Let us now consider the partitioned matrices:
$$
\SIG_k^* = \begin{bmatrix}
			      \overbar{\SIG}_k & \CC_k\\
			      \CC^{'}_k & \SIG^{\text{\tiny Q}}_k
			   \end{bmatrix}, \qquad\quad
\mathbf{O}_k = \begin{bmatrix}
			      \mathbf{W}_k & \mathbf{V}_k\\
			      \mathbf{V}^{'}_k & \mathbf{U}_k
			   \end{bmatrix}.
$$
Furthermore, define $\mathbf{E}_k = \SIG^{\text{\tiny Q}}_k - \CC^{'}_k \overbar{\SIG}_k^{-1} \CC_k$. Then
$$
(\SIG_k^*)^{-1} = \begin{bmatrix}
			      \overbar{\SIG}_k^{-1} + \overbar{\SIG}_k^{-1}\CC_k \mathbf{E}_k^{-1}\CC^{'}_k\overbar{\SIG}_k^{-1}  & -\overbar{\SIG}_k^{-1} \CC_k \mathbf{E}_k^{-1} \\
			      \mathbf{E}_k^{-1}\CC^{'}_k \overbar{\SIG}_k^{-1} & \mathbf{E}_k^{-1}
			   \end{bmatrix},
$$
and $\log\det\SIG_k^* = \log\det\overbar{\SIG}_k + \log\det\mathbf{E}_k$. It follows that $F(\SIG^*)$ can be re-expressed as function of $\mathbf{E}_k$ and $\mathbf{C}_k$ as follows:
\[
\begin{split}
F(\mathbf{E}, \CC) &= \sum_k \text{tr} \lbrace \mathbf{W}_k \overbar{\SIG}_k^{-1} \CC_k \mathbf{E}_k^{-1} \CC_k^{'} \overbar{\SIG}_k^{-1} \rbrace -
2\sum_k \text{tr} \{\mathbf{V}_k^{'} \overbar{\SIG}_k^{-1} \CC_k \mathbf{E}_k^{-1} \}\\
& + \sum_k \text{tr} \{\mathbf{U}_k  \mathbf{E}_k^{-1} \} + 
\sum_k N_k \log\det\mathbf{E}_k + \text{const.} \\
\end{split}
\]
Maximization of $F(\mathbf{E}, \CC)$ with respect to $\mathbf{E}_k$ and $\mathbf{C}_k$ leads to:
$$
\widehat{\CC}_k = ( \overbar{\SIG}^{-1}_k \, \W_k \, \overbar{\SIG}^{-1}_k )^{-1} ( \overbar{\SIG}^{-1}_k \mathbf{V}_k),
$$
$$
\widehat{\mathbf{E}}_k = \dfrac{1}{N_k} \left[ \widehat{\CC}^{'}_k \, \overbar{\SIG}^{-1}_k \, \W_k \, \overbar{\SIG}^{-1}_k \, \widehat{\CC}_k - 
2 \mathbf{V}^{'}_k \overbar{\SIG}^{-1}_k \, \widehat{\CC}_k + \mathbf{U}_k \right].
$$
Consequently we have that:
$$ 
\widehat{\SIG}^{\text{\tiny Q}}_k = \widehat{\mathbf{E}}_k + \widehat{\CC}_k^{'} \, \overbar{\SIG}^{-1}_k \, \widehat{\CC}_k.
$$

Given estimates $\widehat{\CC}_k$ and $\widehat{\mathbf{E}}_k$, for the mean parameter $\MU^{\text{\tiny Q}}_k$ corresponding to the additional variables, define now $\mathbf{m}_{ik} = \MU^{\text{\tiny Q}}_k + \CC_k^{'} \overbar{\SIG}^{-1}_k (\yp_i - \overbar{\MU}_k )$. Consequently, the function $F(\MU^*, \SIG^*)$ can be rewritten as:
$$
F(\mathbf{m}) = \sumN \left[ \, \sum_{k=1}^K t_{ik}\, \text{log} \Big\lbrace \, \phi ( \yq_i \vbar \yp_i \,;\, \mathbf{m}_{ik}, \widehat{\mathbf{E}}_k  )  \Big\rbrace \right] + \text{const.}
$$
By plugging the $\mathbf{m}_{ik}$ expression above in $F(\mathbf{m})$, we can express the latter in terms of $\MU^{\text{\tiny Q}}_k$ as:
\[
\begin{split}
F(\MU^{\text{\tiny Q}}) &= -\frac{1}{2} \sumN \sum_{k=1}^K t_{ik} \Bigl\{ \bigl[ \yq_i - \MU^{\text{\tiny Q}}_k -  \widehat{\CC}_k^{'} \overbar{\SIG}_k^{-1}(\yp_i -\overbar{\MU}_k) \bigr]^{'} \widehat{\mathbf{E}}^{-1}_k \bigl[\y_i -\MU^{\text{\tiny Q}}_k -  \widehat{\CC}_k^{'} \overbar{\SIG}_k^{-1}(\y^{\text{\tiny P}}_i -\overbar{\MU}_k)\bigr] \Bigr\}\\
&+ \text{const}.
\end{split}
\]
Taking derivatives of $F(\MU^{\text{\tiny Q}})$ and solving for $\MU^{\text{\tiny Q}}_k$ we obtain:
$$
\widehat{\MU}^{\text{\tiny Q}}_k = \dfrac{1}{N_k} \left[\, \sumN t_{ik} \yq_i - \widehat{\CC}^{'}_k \, \overbar{\SIG}^{-1}_k \sumN t_{ik} (\yp_i - \overbar{\MU}_k ) \right].
$$
The above passages prove the derivation of the updating equations of the M step in Section \ref{mstep_ice}.

\section{A note on regularization}\label{app:regularization}
The procedure described in \ref{mstep_ice} requires the empirical class scatter matrix $\OO_k$ to be definite positive. This may not be the case in situations where the expected number of observations in a class is small or the variables are highly correlated. Approaches for Bayesian regularization in the context of finite Gaussian mixture models for clustering have already been suggested in the literature, see in particular \cite{baudry:2015}. We suggest a similar approach, proposing the following regularized version of $\OO_k$:
$$
\OO^{\text{\tiny{reg}}}_k = \OO_k + \dfrac{\mathbf{S}}{\det(\mathbf{S})^{1/R}}\left( \dfrac{\gamma}{K+H} \right)^{1/R},
$$
where $\mathbf{S} = \frac{1}{N}\sumN ( \y_i - \bar{\y} )( \y_i - \bar{\y} )^{'}$ is the empirical covariance matrix computed on the full test data, and $\bar{\y}$ the sample mean, $\bar{\y} = \frac{1}{N}\sumN \y_i$. The coefficient $\gamma$ controls the amount of regularization and we set it to $(\log R) / N$; see \cite{baudry:2015} for further details.

\section{Details and results of simulation experiments}
\label{app:sim}
In this section we describe in more details the settings of the simulated data experiments and we give a graphical summary of the results. 

\subsection{Simulation study 1}
The training data has $M = 300$ observations in all scenarios. A random subset of the 27 variables is taken from the data with the number of variables observed in the training set equal $P = \{18, 9, 3\}$. Test set on all the 27 variables and different sample sizes equal to $N = \{50,100,200,300,500\}$. Different scenarios are defined by different combinations of $P$ and $N$, One class is randomly deleted from the training data, while all 3 classes are present in the test data. In each scenario, the following models are considered: \textsf{full}, the EDDA classifier fitted on the training data with full information, i.e. all 3 classes and all 27 variables, tested on the full test data; \textsf{EDDA} the EDDA classifier fitted on the training data considering only a subset of the variables, then tested on the full test data; the \textsf{AMDA} approach of \cite{bouveyron:2014} fitted on the simulated training data with a subset of the variables and tested on the test data with the subset of variables observed in the training; the presented \textsf{D-AMDA} framework. Each experiment is replicated 100 times for all combinations of sample sizes and number of observed training variables. Model selection for AMDA and D-AMDA is performed using BIC and a range of values of $H$ from 0 to 4. 
Since AMDA and D-AMDA are partially unsupervised, we compute the classification error on the matching classes detected after tabulating the actual classification with the estimated one using function \texttt{matchClasses} of package \texttt{e1071} \citep{e1071}. To compare AMDA and D-AMDA, we also report the adjusted Rand index \citep[ARI,][]{hubert:arabie:1985}. Indeed, the learning in the test set is partly unsupervised, and a number of hidden classes different from 1 could be estimated.

The next figures contain graphical summaries of the results of the various experiments for the different scenarios.

\begin{figure}[H]
 \centering
 \includegraphics[scale=0.57]{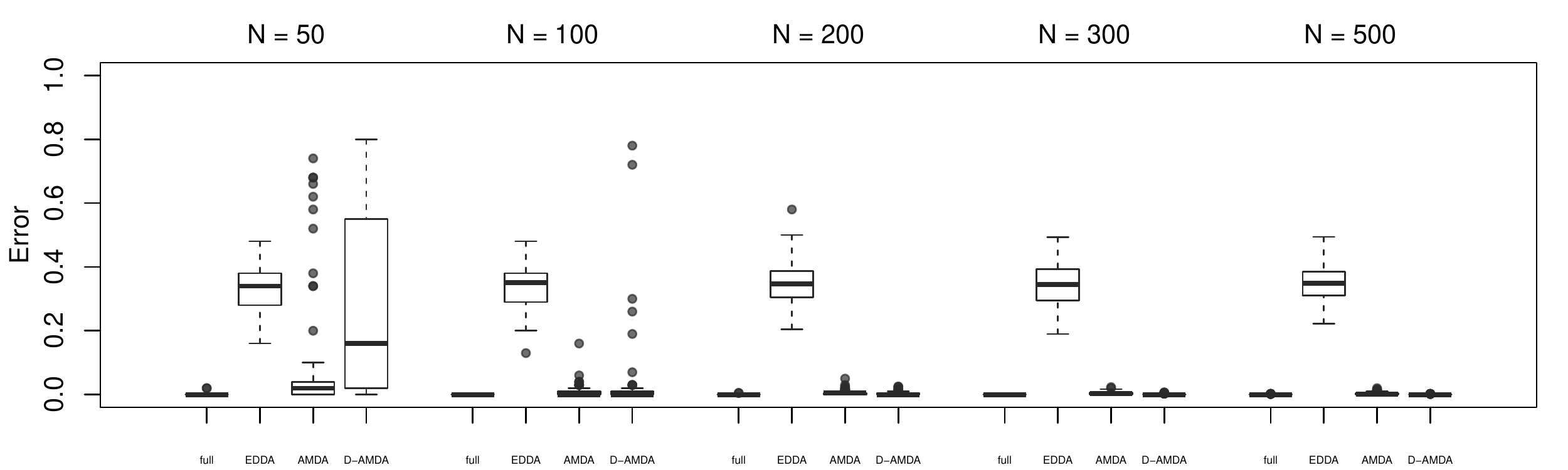}
 \caption{\label{fig_sim_wine_P1_err} Simulation study 1, $P = 18$. Error computed on the matched classes between the actual classification of the test data and the estimated one.}
\end{figure}
\begin{figure}[H]
 \centering
 \includegraphics[scale=0.57]{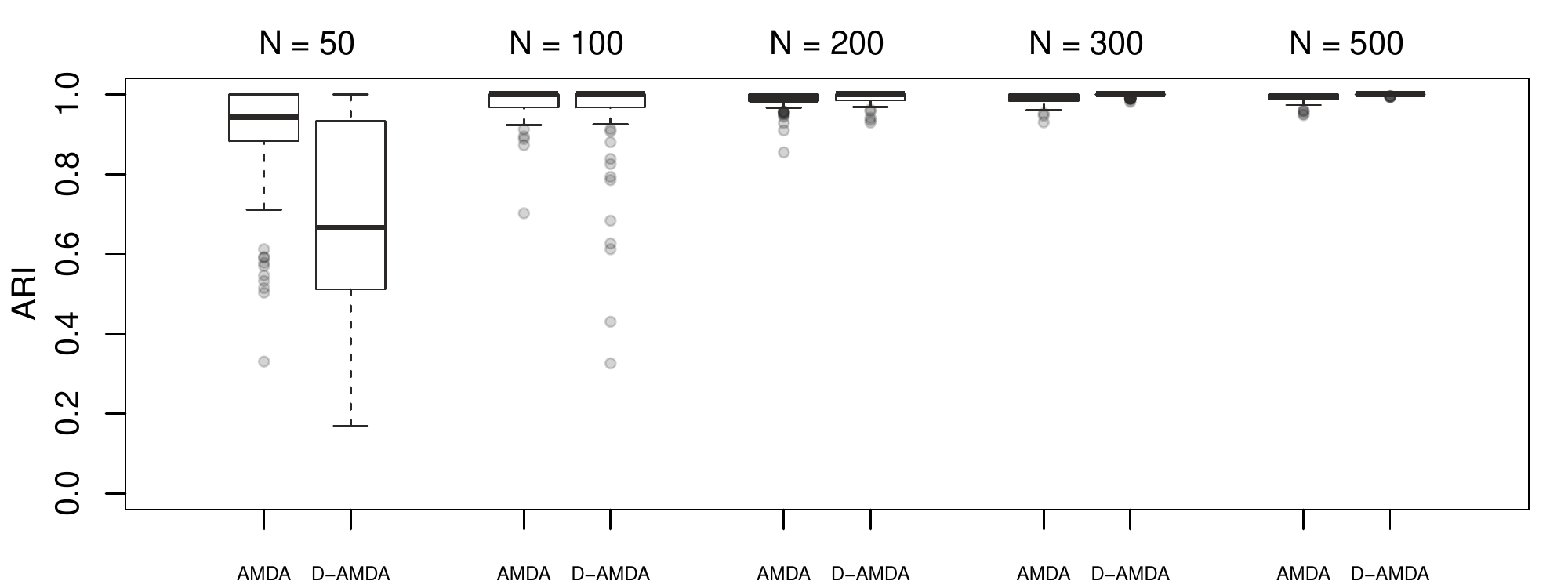}
 \caption{\label{fig_sim_wine_P1_ari} Simulation study 1, $P = 18$. Adjusted Rand index between the actual classification of the test data and the estimated one for AMDA and D-AMDA.}
\end{figure}

\begin{figure}[H]
 \centering
 \includegraphics[scale=0.57]{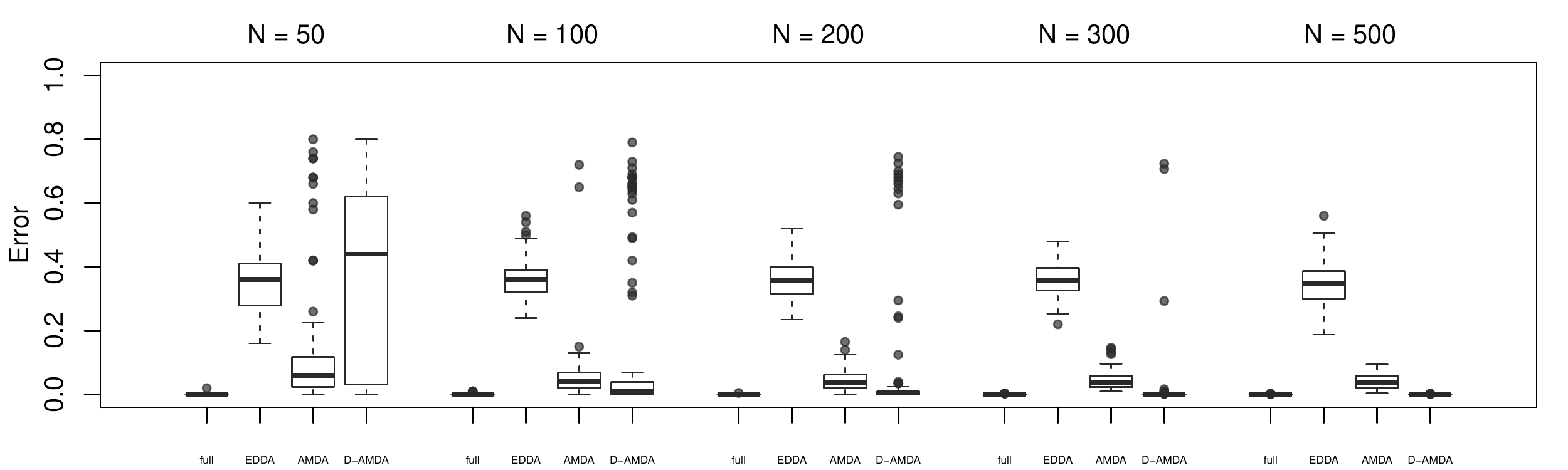}
 \caption{\label{fig_sim_wine_P2_err} Simulation study 1, $P = 9$. Error computed on the matched classes between the actual classification of the test data and the estimated one.}
\end{figure}
\begin{figure}[H]
 \centering
 \includegraphics[scale=0.57]{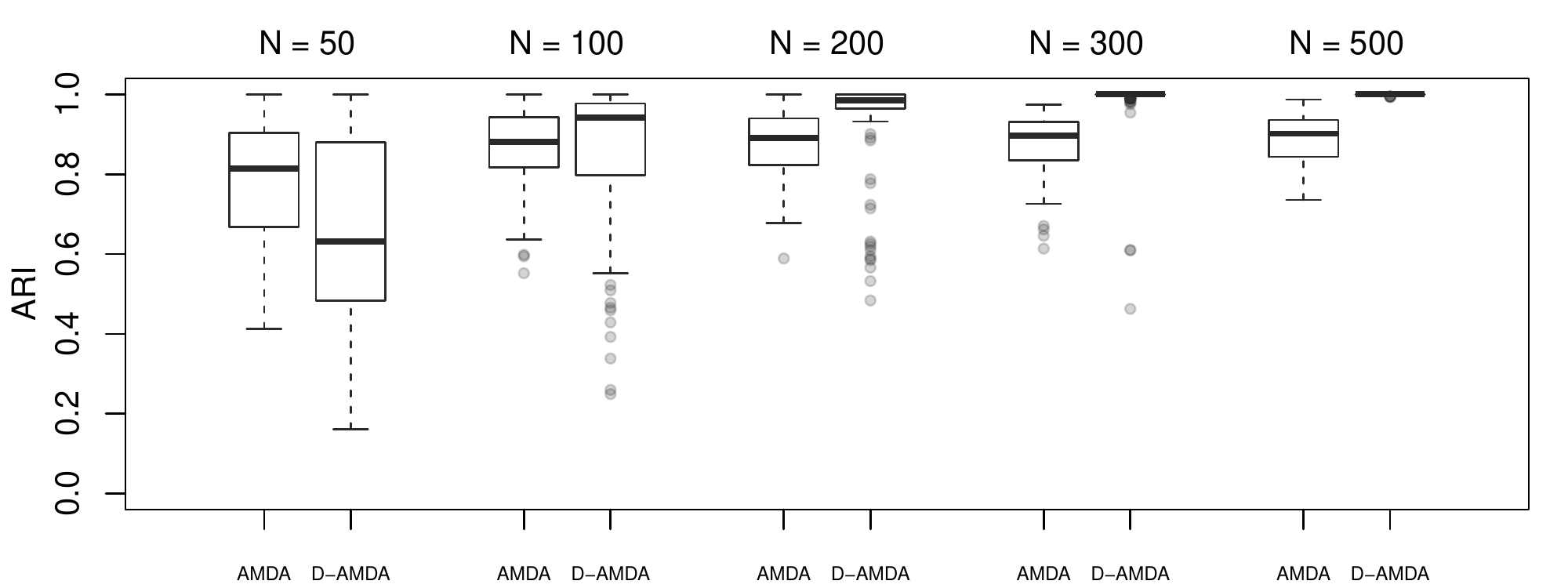}
 \caption{\label{fig_sim_wine_P2_ari} Simulation study 1, $P = 9$. Adjusted Rand index between the actual classification of the test data and the estimated one for AMDA and D-AMDA.}
\end{figure}

\begin{figure}[H]
 \centering
 \includegraphics[scale=0.57]{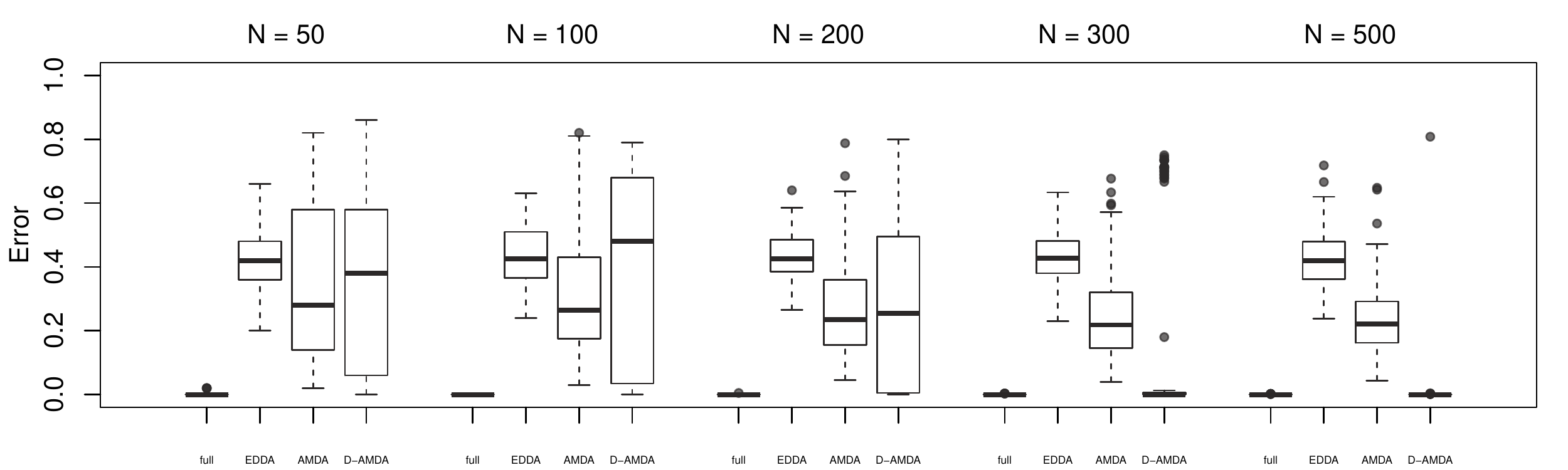}
 \caption{\label{fig_sim_wine_P3_err} Simulation study 1, $P = 3$. Error computed on the matched classes between the actual classification of the test data and the estimated one.}
\end{figure}
\begin{figure}[H]
 \centering
 \includegraphics[scale=0.57]{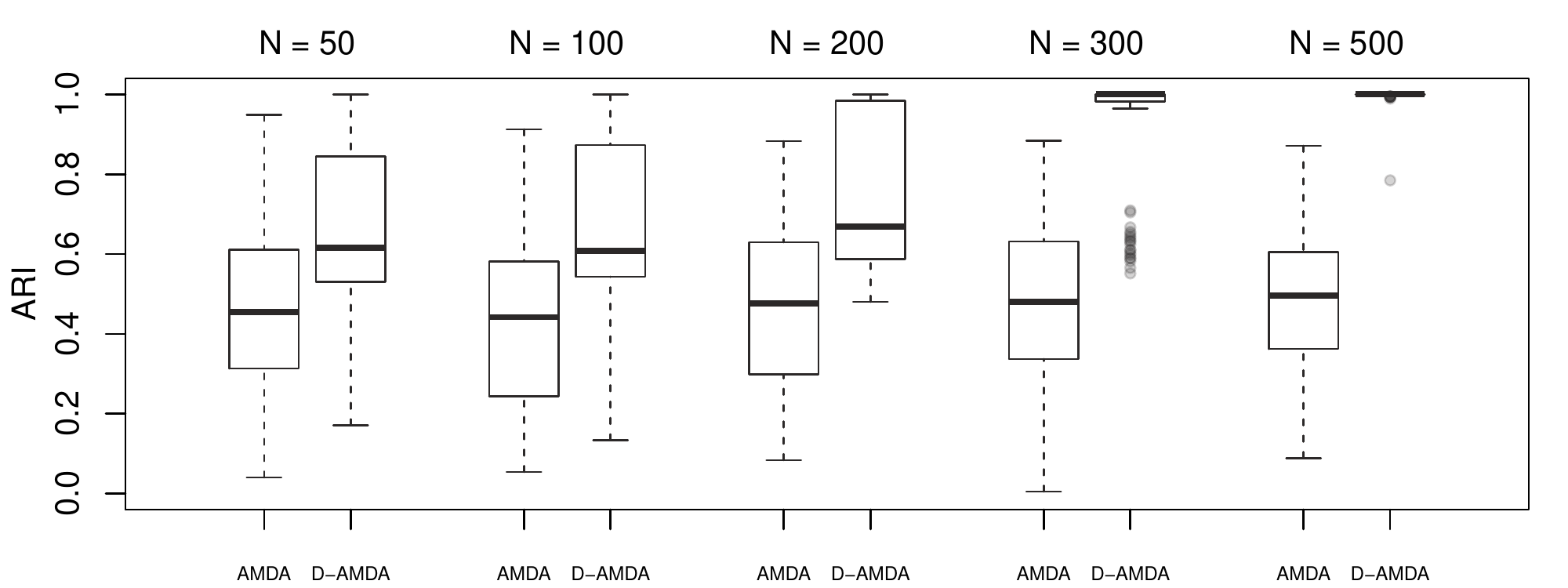}
 \caption{\label{fig_sim_wine_P3_ari} Simulation study 1, $P = 3$. Adjusted Rand index between the actual classification of the test data and the estimated one for AMDA and D-AMDA.}
\end{figure}

\subsection{Simulation study 2}
\Gen{} variables are distributed according to a mixture of $C=4$ multivariate Gaussian distributions with mixing proportions (0.3, 0.4, 0.4, 0.3). Mean parameters are randomly chosen in (-7, 7), (-4.5, 4.5), (-0.5, 0.5), (-10, 10). For each class, the covariance matrices are randomly generated from the Wishart distributions $\mathcal{W}(G, \Psi_1)$, $\mathcal{W}(G + 2, \Psi_2)$, $\mathcal{W}(G + 1, \Psi_3)$, $\mathcal{W}(G, \Psi_4)$, where $G$ denotes the number of generative variables. The scale matrices are respectively defined: $\Psi_1$, is such that $\psi_{jj} = 1$ and $\psi_{ji} = \psi_{ij} = 0.7$; $\Psi_3$, is such that $\psi_{jj} = 1$ and $\psi_{ji} = \psi_{ij} = 0.5$; $\Psi_2 = \Psi_4 = \I$. \Cor{} variables are generated as $X_{g_1} + X_{g_2} + \epsilon$, where $X_{g_1}$ and $X_{g_2}$ are two randomly chosen \Gen{} variables and $\epsilon \sim \mathcal{N}(0,1)$. In Simulations 1 and 2, \Noi{} variables are generated as $\mathcal{N}(\mathbf{0}, \Psi)$, where $\Psi$ is such that $\psi_{jj} = 1$ and $\psi_{ji} = \psi_{ij} = 0.5$; thus they are correlated to each other, but not to \Cor{} and \Gen{} variables. In Simulation 3, the \Noi{} variables are generated all independent of each other.  The 2 classes observed in the learning set are randomly chosen from the set of 4 classes with equal probabilities. We considered three different sample sizes for the test data, respectively 100, 200 and 400. Each scenario within each experiment and for each sample size was replicated 50 times.

The next subsections contain graphical summaries of the various experiments for the different scenarios. Throughout the different scenarios, we used the ARI to assess the quality of the classification and we compared the results of the following methods:
\begin{itemize}[noitemsep]
 \item \textsf{noSel}, the D-AMDA model applied on $\X$ and the full $\Y$ without performing any variable selection.
 \item \textsf{orac}, representing the ``oracle'' solution. This corresponds to the D-AMDA model applied on $\X^*$ and $\Y^*$, the learning and test set where only \Gen{} variables are observed.
 \item \textsf{varSel}, the D-AMDA model with the forward variable selection applied to the observed $\X$ and $\Y$. 
\end{itemize}
The variable selection performance of the proposed method was assessed via the proportion of times each variable was selected as relevant out of the 50 replicated experiments.

\subsubsection{Experiment 1}
\begin{figure}[H]
 \centering
 \includegraphics[scale=0.57]{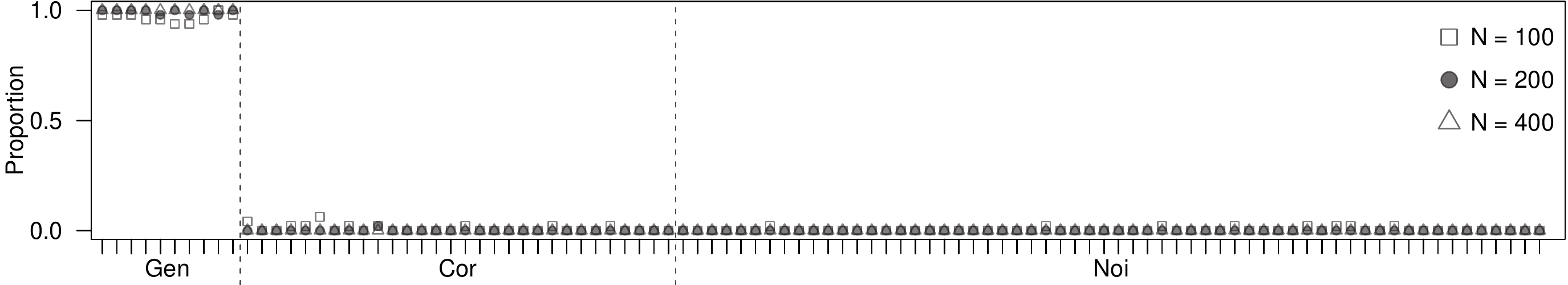}
 \caption{\label{fig_sim1_s1} Simulation experiment 1, scenario (a). Proportions of time a variable has been selected as relevant. }
\end{figure}
\begin{figure}[H]
 \centering
 \includegraphics[scale=0.57]{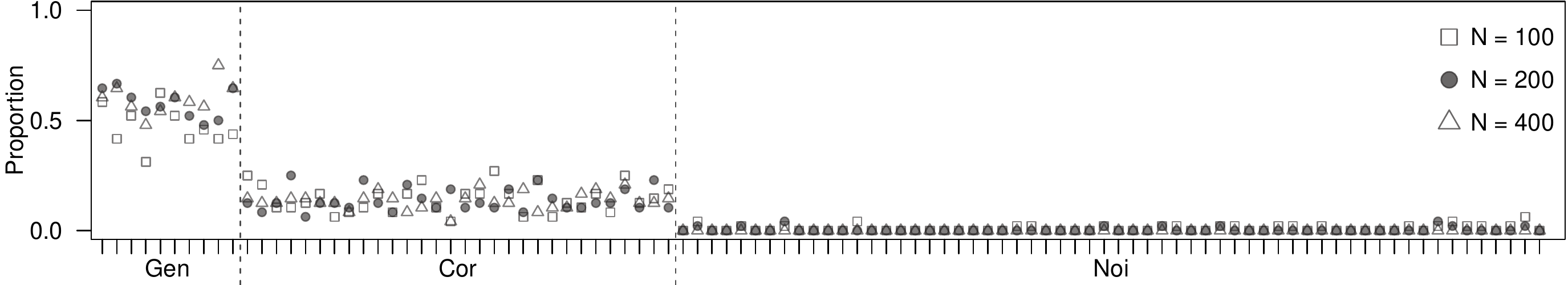}
 \caption{\label{fig_sim1_s2} Simulation experiment 1, scenario (b). Proportions of time a variable has been selected as relevant.}
\end{figure}
\begin{figure}[H]
 \centering
 \includegraphics[scale=0.57]{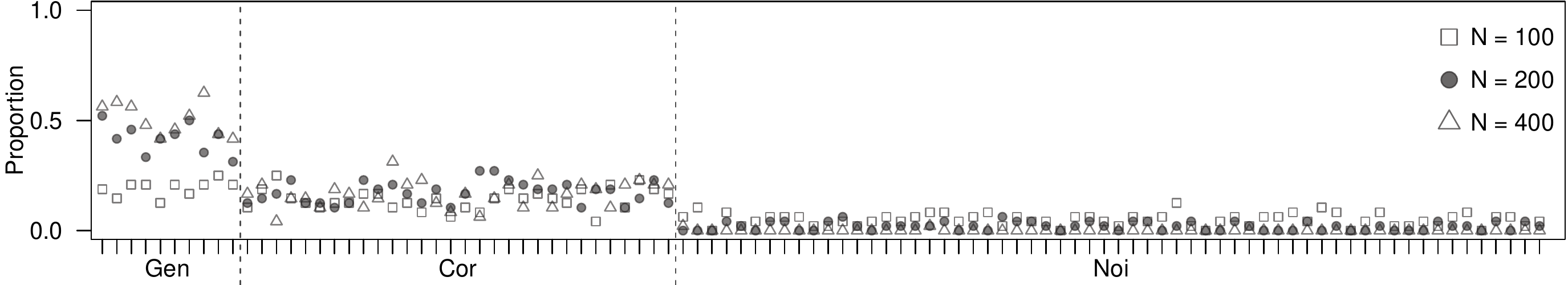}
 \caption{\label{fig_sim1_s3} Simulation experiment 1, scenario (c). Proportions of time a variable has been selected as relevant.}
\end{figure}

\begin{figure}[H]
 \centering
 \includegraphics[scale=0.57]{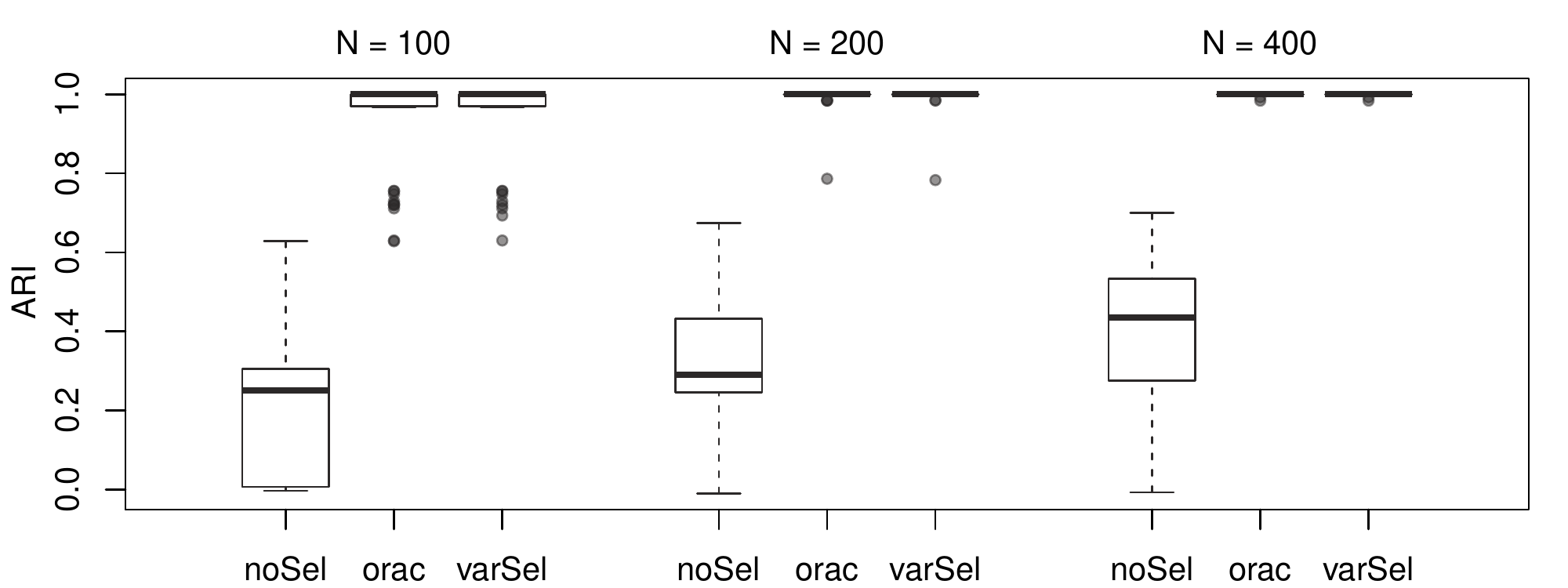}
 \caption{\label{fig_sim1_s1_ari} Simulation experiment 1, scenario (a). ARI between the actual classification of the test data and the estimated one.}
\end{figure}
\begin{figure}[H]
 \centering
 \includegraphics[scale=0.57]{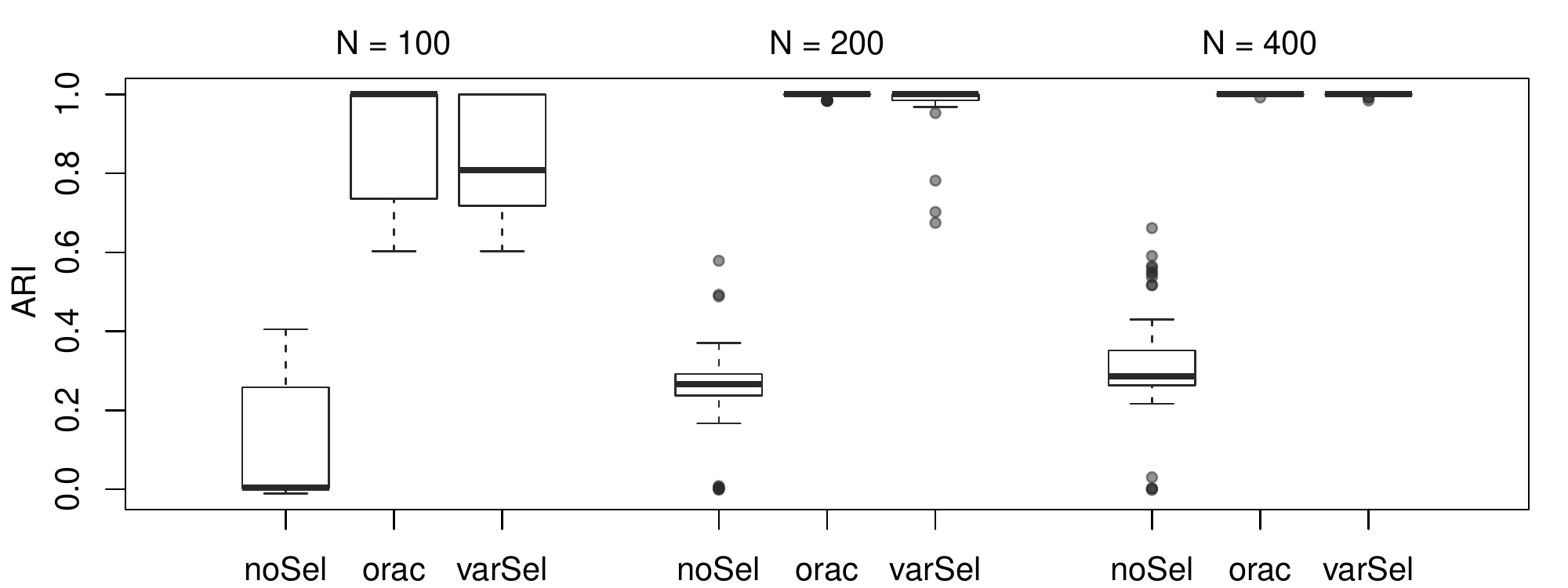}
 \caption{\label{fig_sim1_s2_ari} Simulation experiment 1, scenario (b). ARI between the actual classification of the test data and the estimated one.}
\end{figure}
\begin{figure}[H]
 \centering
 \includegraphics[scale=0.57]{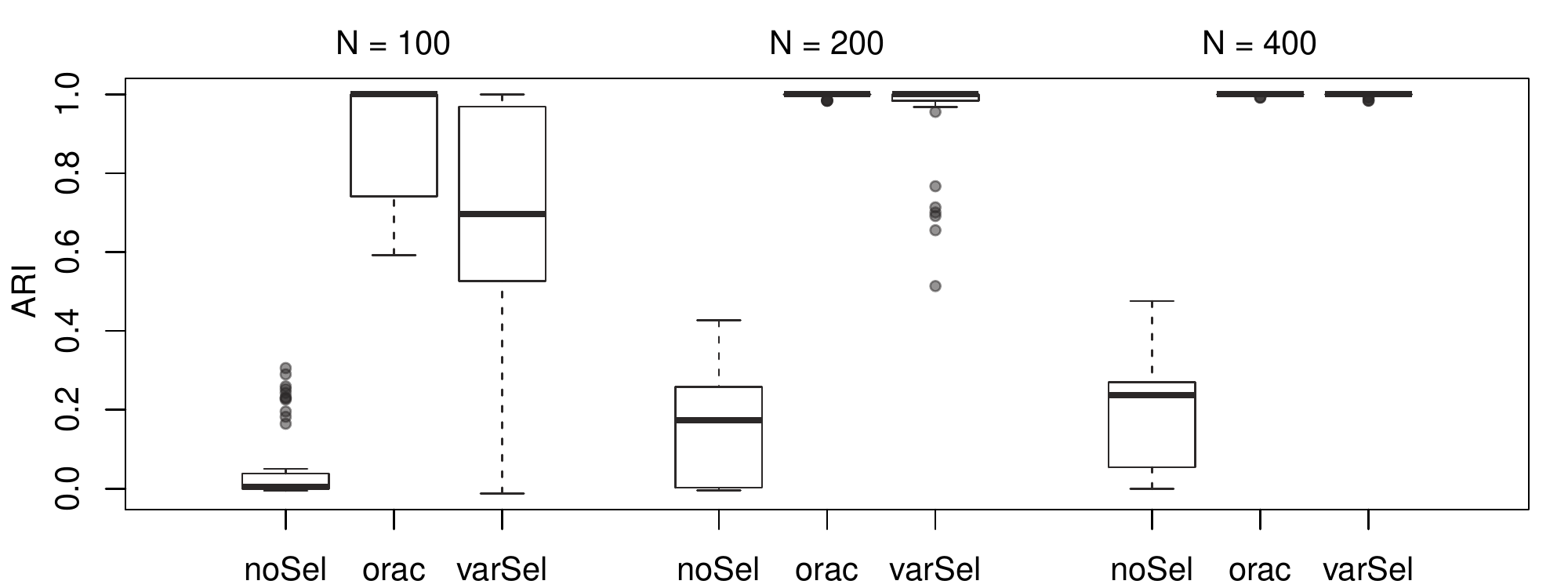}
 \caption{\label{fig_sim1_s3_ari} Simulation experiment 1, scenario (c). ARI between the actual classification of the test data and the estimated one.}
\end{figure}

\subsubsection{Experiment 2}
\begin{figure}[H]
 \centering
 \includegraphics[scale=0.57]{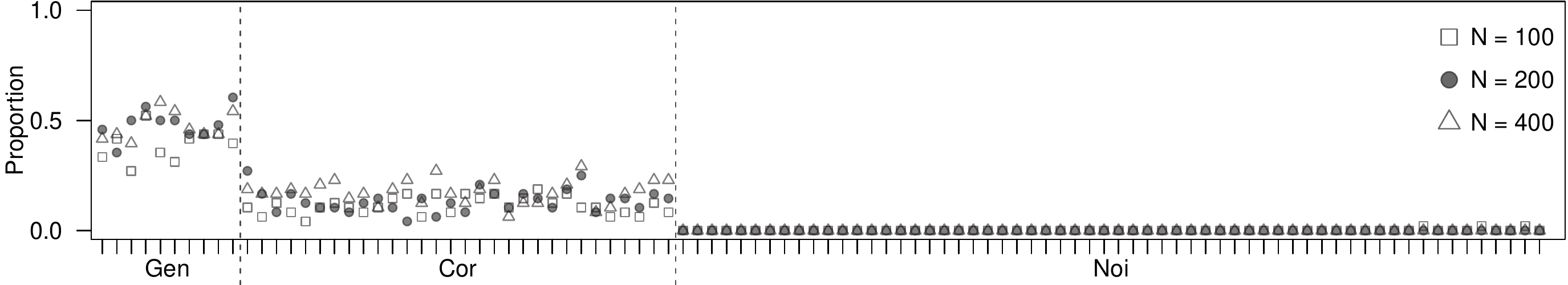}
 \caption{\label{fig_sim2_s1} Simulation experiment 2, scenario (a). Proportions of time a variable has been selected as relevant.}
\end{figure}
\begin{figure}[H]
 \centering
 \includegraphics[scale=0.57]{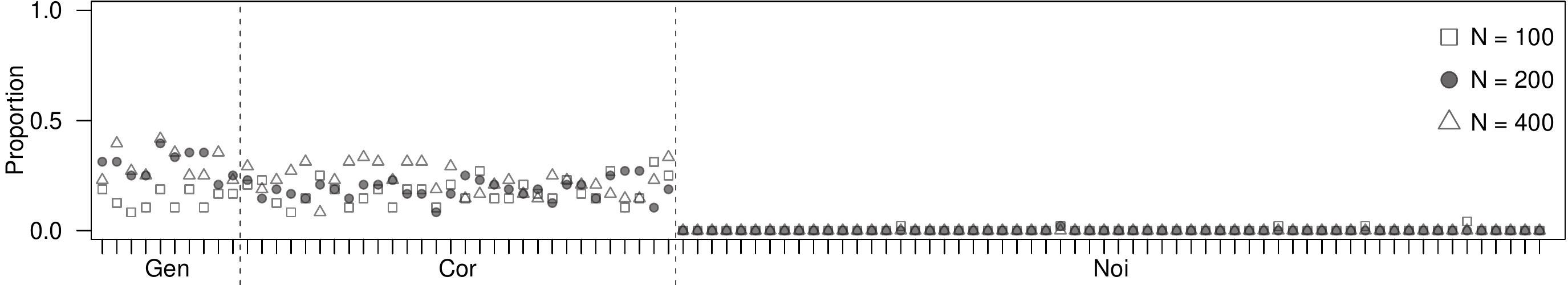}
 \caption{\label{fig_sim2_s2} Simulation experiment 2, scenario (b). Proportions of time a variable has been selected as relevant.}
\end{figure}
\begin{figure}[H]
 \centering
 \includegraphics[scale=0.57]{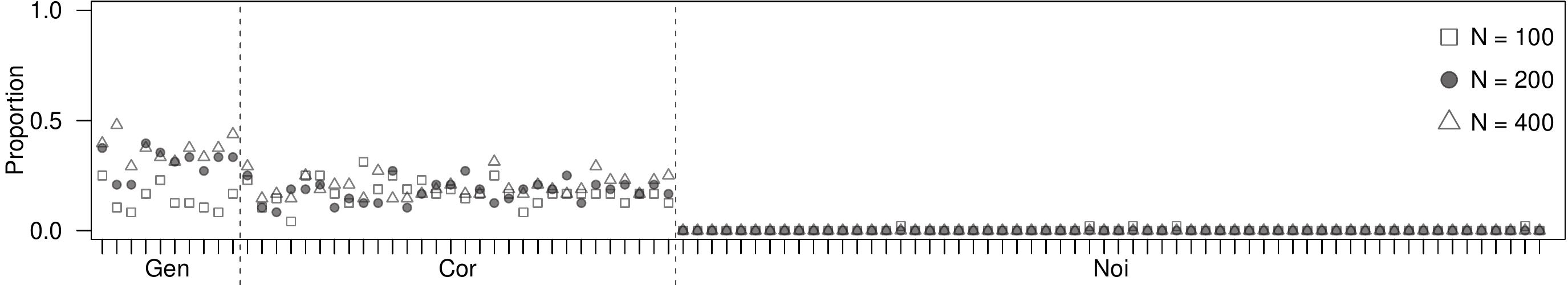}
 \caption{\label{fig_sim2_s3} Simulation experiment 2, scenario (c). Proportions of time a variable has been selected as relevant.}
\end{figure}

\begin{figure}[H]
 \centering
 \includegraphics[scale=0.57]{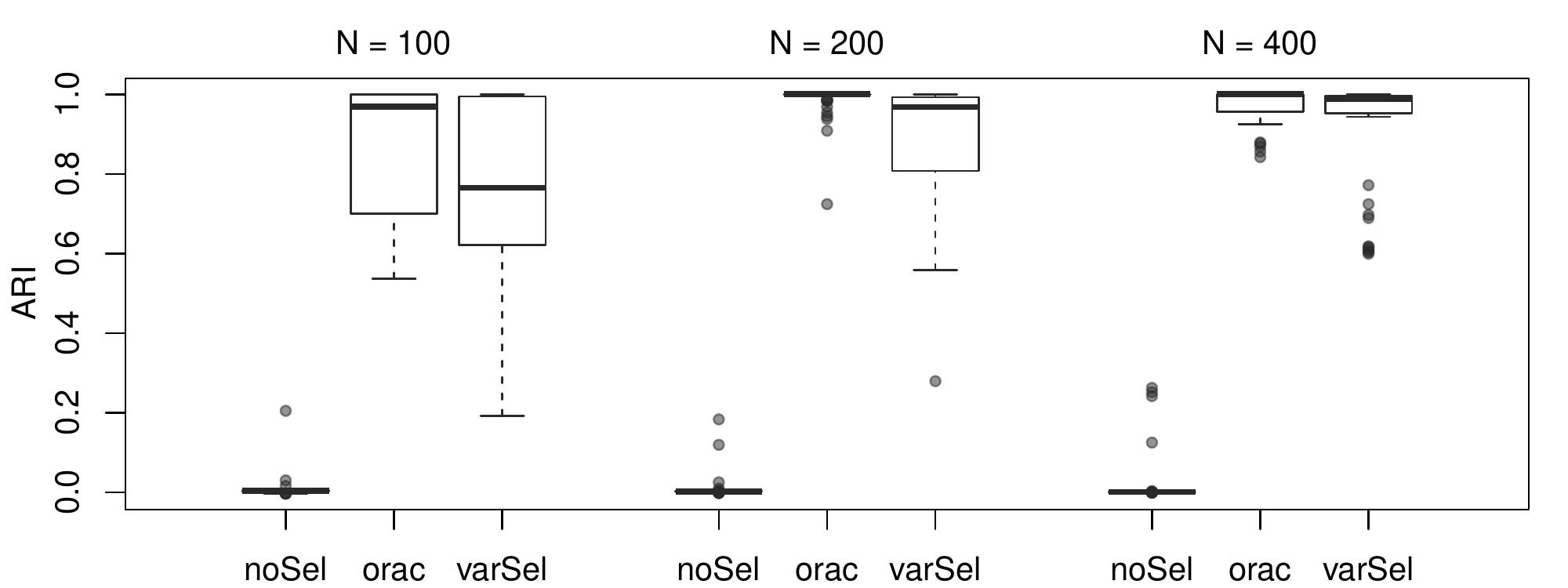}
 \caption{\label{fig_sim2_s1_ari} Simulation experiment 2, scenario (a). ARI between the actual classification of the test data and the estimated one.}
\end{figure}
\begin{figure}[H]
 \centering
 \includegraphics[scale=0.57]{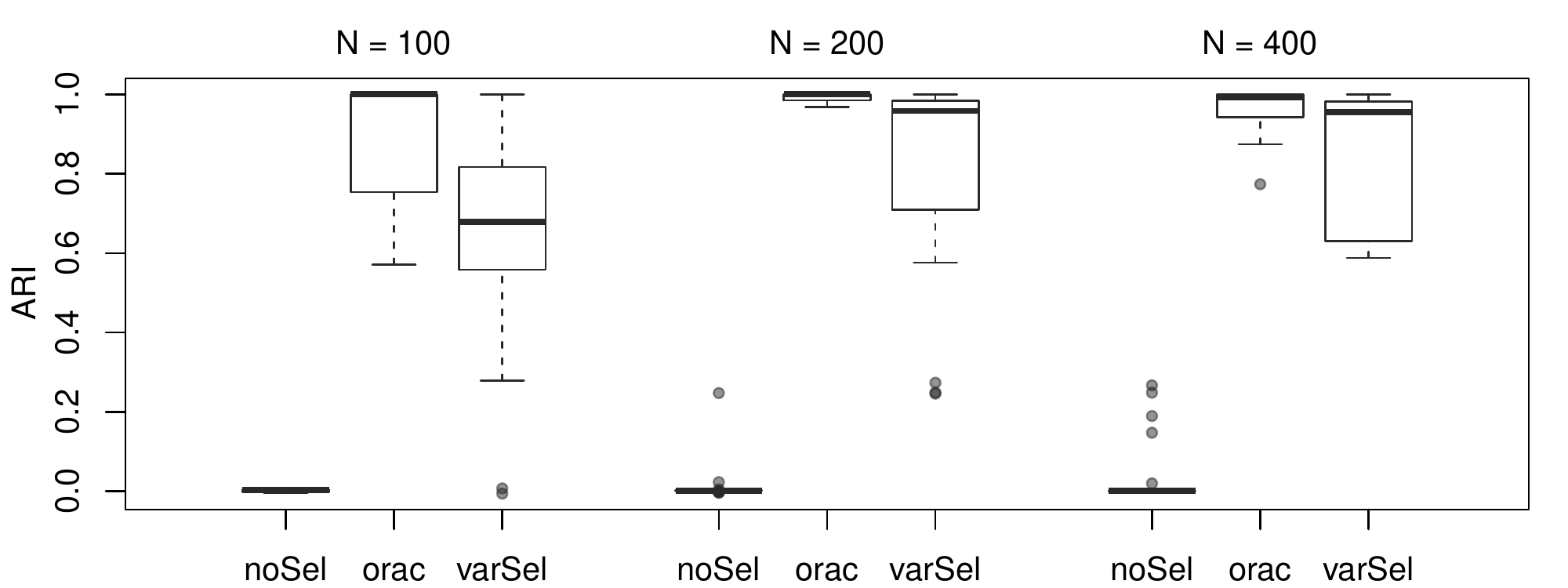}
 \caption{\label{fig_sim2_s2_ari} Simulation experiment 2, scenario (b). ARI between the actual classification of the test data and the estimated one.}
\end{figure}
\begin{figure}[H]
 \centering
 \includegraphics[scale=0.57]{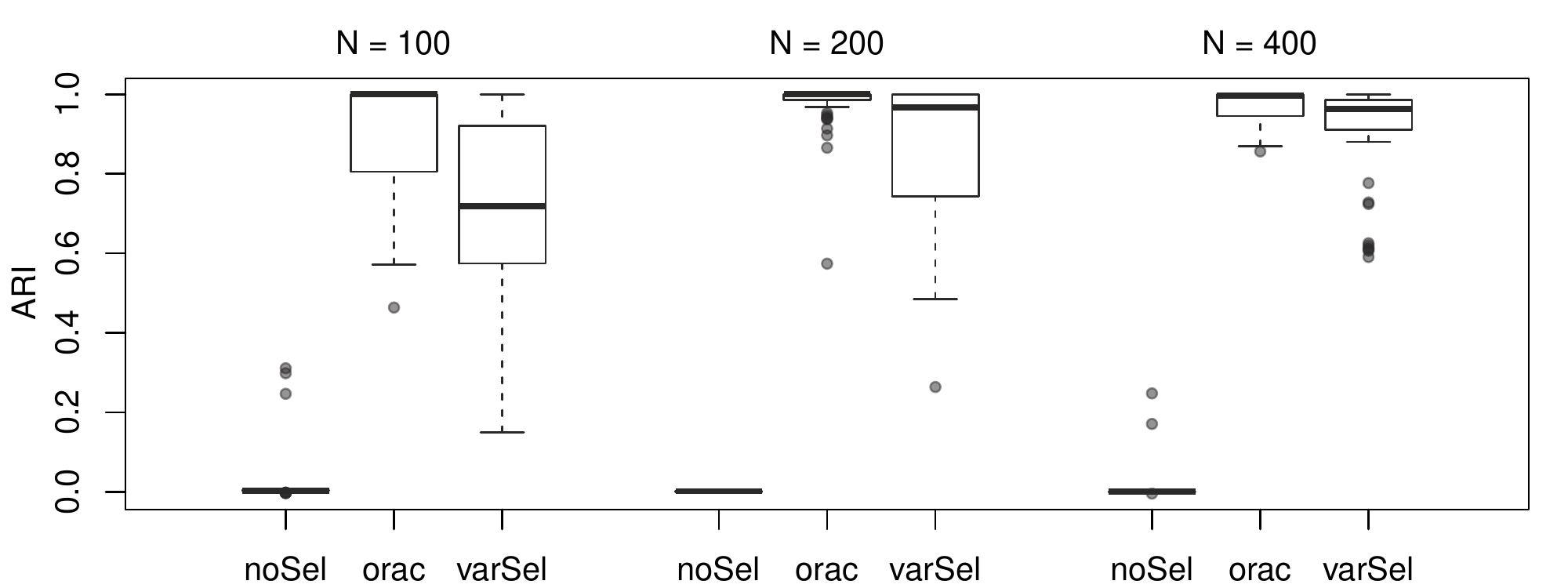}
 \caption{\label{fig_sim2_s3_ari} Simulation experiment 2, scenario (c). ARI between the actual classification of the test data and the estimated one.}
\end{figure}

\subsubsection{Experiment 3}
\begin{figure}[H]
 \centering
 \includegraphics[scale=0.57]{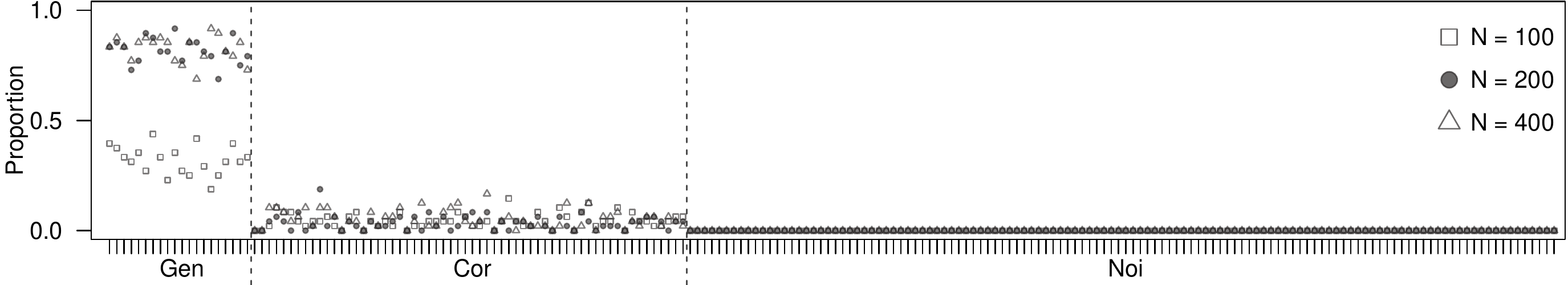}
 \caption{\label{fig_sim3_s1} Simulation experiment 3, scenario (a). Proportions of time a variable has been selected as relevant.}
\end{figure}
\begin{figure}[H]
 \centering
 \includegraphics[scale=0.57]{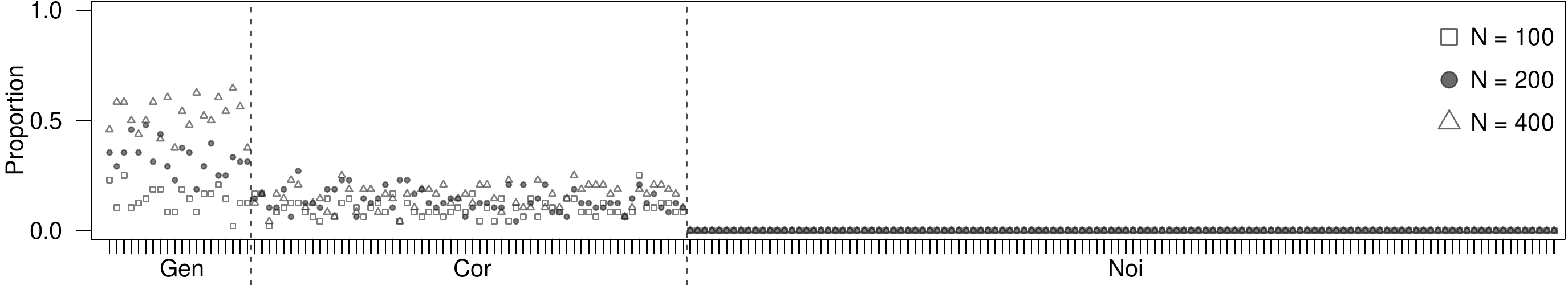}
 \caption{\label{fig_sim3_s2} Simulation experiment 3, scenario (b). Proportions of time a variable has been selected as relevant.}
\end{figure}
\begin{figure}[H]
 \centering
 \includegraphics[scale=0.57]{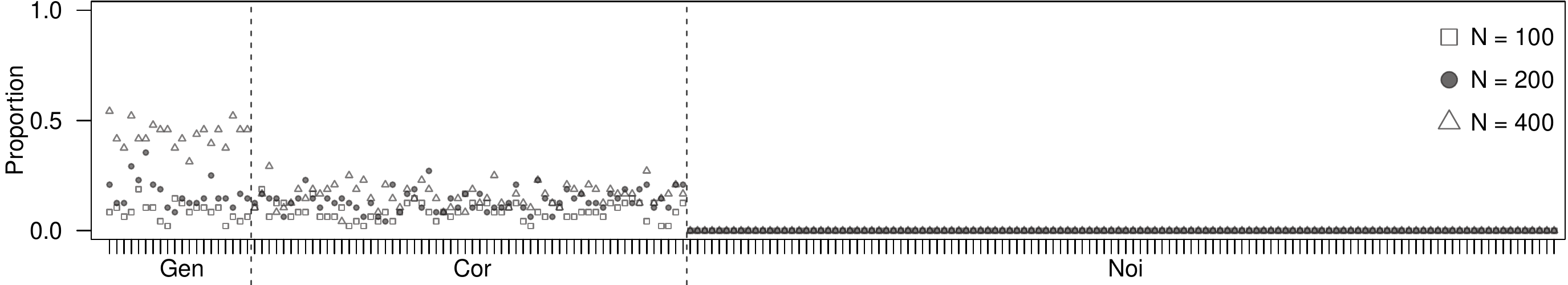}
 \caption{\label{fig_sim3_s3} Simulation experiment 3, scenario (c). Proportions of time a variable has been selected as relevant.}
\end{figure}

\begin{figure}[H]
 \centering
 \includegraphics[scale=0.57]{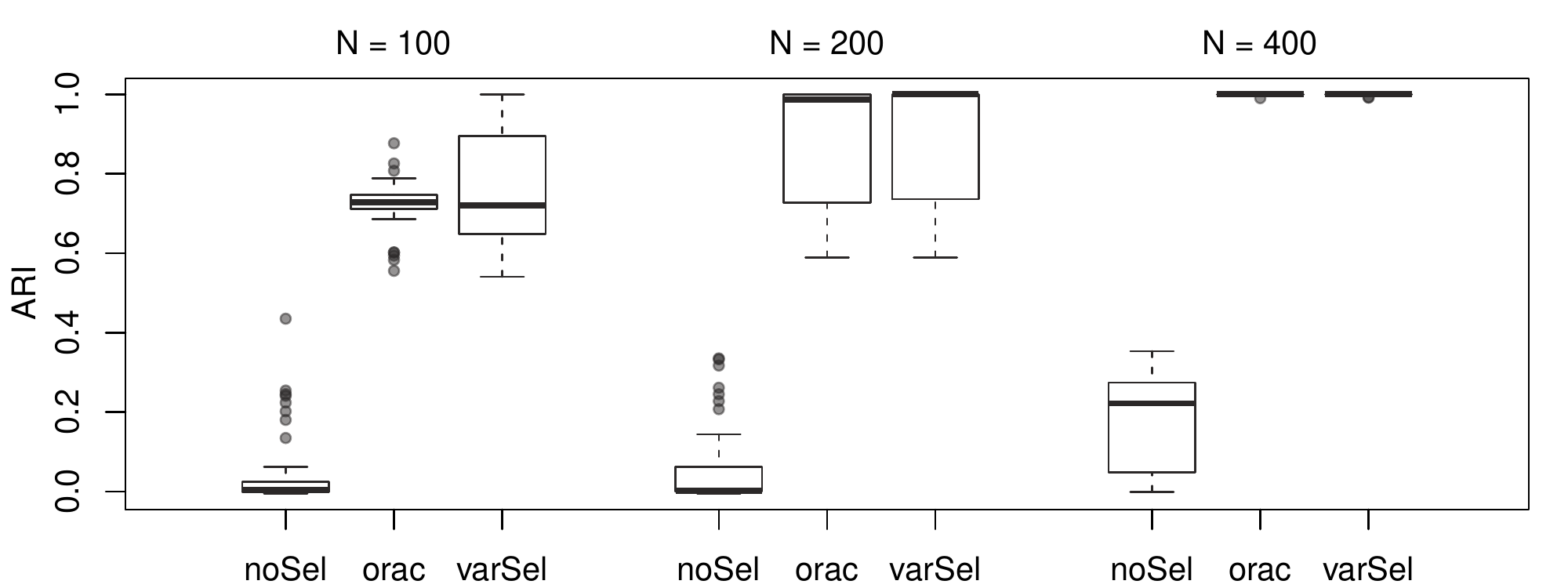}
 \caption{\label{fig_sim3_s1_ari} Simulation experiment 3, scenario (a). ARI between the actual classification of the test data and the estimated one.}
\end{figure}
\begin{figure}[H]
 \centering
 \includegraphics[scale=0.57]{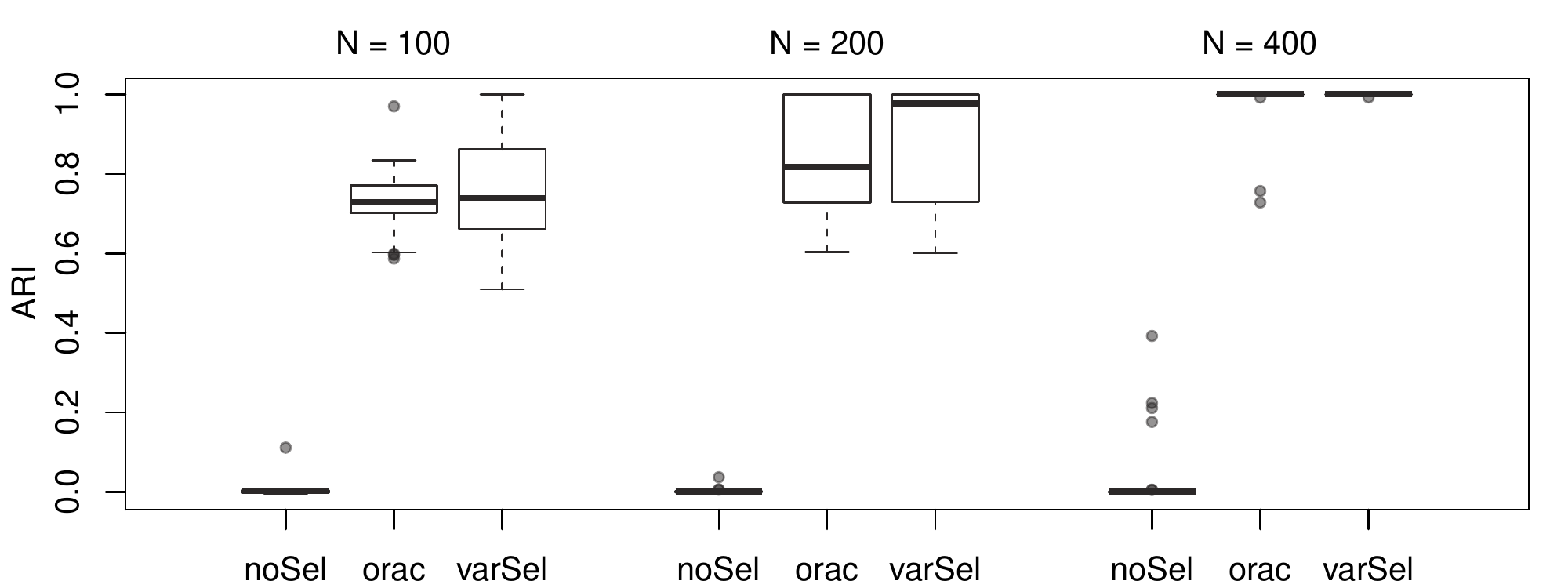}
 \caption{\label{fig_sim3_s2_ari} Simulation experiment 3, scenario (b). ARI between the actual classification of the test data and the estimated one.}
\end{figure}
\begin{figure}[H]
 \centering
 \includegraphics[scale=0.57]{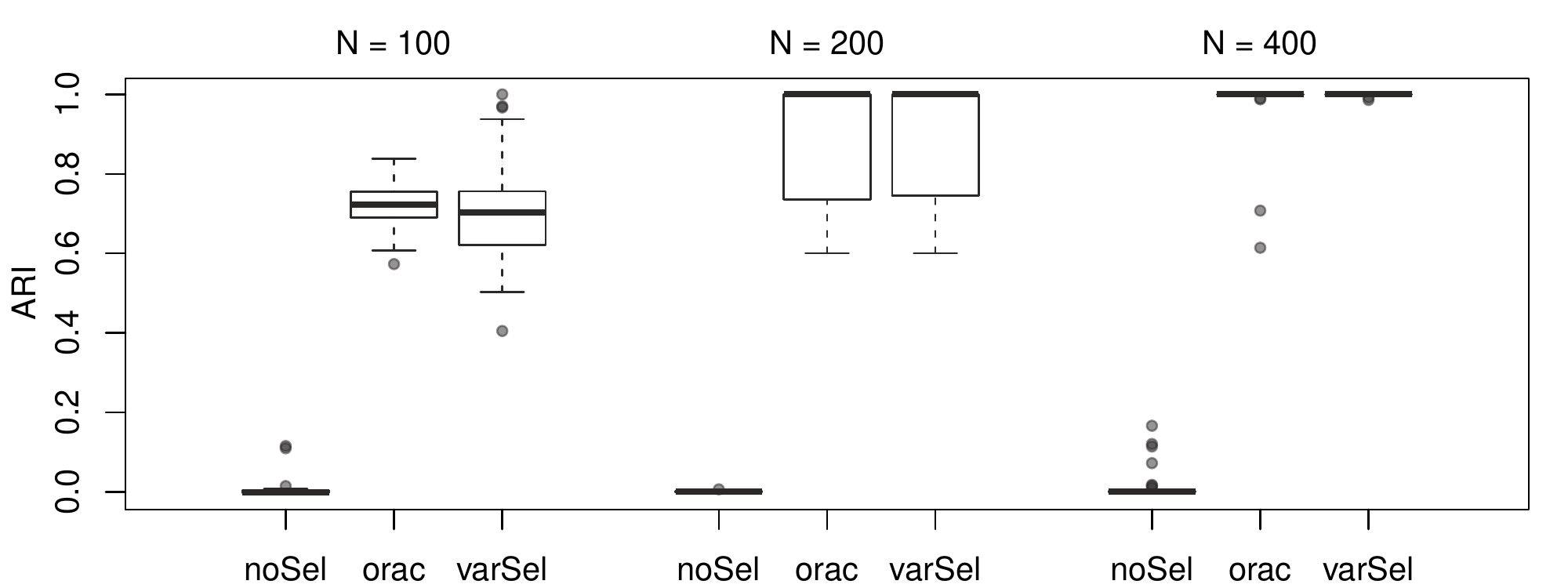}
 \caption{\label{fig_sim3_s3_ari} Simulation experiment 3, scenario (c). ARI between the actual classification of the test data and the estimated one.}
\end{figure}

\newpage
\bibliographystyle{apalike}
\bibliography{bibliography.bib}

\end{document}